\documentclass[a4paper,11pt]{article}
\pdfoutput=1 % if your are submitting a pdflatex (i.e. if you have
\usepackage{jheppub} % for details on the use of the package, please
\usepackage[T1]{fontenc} % if needed

\usepackage{subfigure} % Allows subfigures
\usepackage{booktabs} %Fancy tables: use toprule, midrule, and bottomrule.

\newcommand{\as}{\alpha_s}
\newcommand{\wh}{\widehat}

\newcommand{\ket}[1]{\left|#1\right\rangle}

\newcommand{\gsim}
{\;\raisebox{-.3em}{$\stackrel{\displaystyle >}{\sim}$}\;}

\newcommand{\MSb}{{\overline{\rm MS}}}

\def\beq {\begin{equation}}
\def\eeq {\end{equation}}
\def\bea {\begin{eqnarray}}
\def\eea {\end{eqnarray}}

\def\nn {\nonumber}

\def\Im {\mbox{Im}}
\def\Re {\mbox{Re}}

\preprint{UAB-FT-723, TUM-HEP-863/12}

\title{\boldmath Perturbative expansion of $\tau$ hadronic spectral
 function moments and $\alpha_s$ extractions}
\author[a]{Martin~Beneke,}
\author[a]{Diogo~Boito,}
\author[b]{and Matthias~Jamin}

\affiliation[a]{Physik Department T31, Technische Universit\"at M\"unchen, \\
James-Franck-Stra\ss e 1, D-85748 Garching, Germany}

\affiliation[b]{Instituci\'o Catalana de Recerca i Estudis Avan\c cats (ICREA),\\
IFAE, Theoretical Physics Group, UAB,\\ E-08193 Bellaterra, Barcelona, Spain}

\emailAdd{diogo.boito@tum.de}

\abstract{
Various moments of the hadronic spectral functions have been employed in the
determination of the strong coupling $\alpha_s$ from tau decays. In this work
we study the behaviour of their perturbative series under different assumptions
for the large-order behaviour of the Adler function, extending previous work on
the tau hadronic width. We find that the moments can be divided into a small
number of classes, whose characteristics depend only on generic features of
the moment weight function and Adler function series. Some moments that are commonly employed in $\alpha_s$ analyses from $\tau$ decays should be avoided because of their perturbative instability. This conclusion is corroborated by a simplified $\alpha_s$ extraction from individual moments.  Furthermore, under reasonable assumptions for the higher-order behaviour of the perturbative series, fixed-order perturbation theory (FOPT) provides the preferred framework for the renormalization group improvement of all moments that show good perturbative behaviour.
Finally, we provide further evidence for the
plausibility of the description of the Adler function in terms of a small
number of leading renormalon singularities.}

\keywords{$\tau$ decays, $\alpha_s$}

\begin{document}
\maketitle
\flushbottom
\section{Introduction}

The precise determination of fundamental parameters of the Standard Model (SM)
provides one of the most important tests of its internal consistency. In the
strong sector, the QCD coupling $\alpha_s$ plays a prominent role and much
effort has been devoted to its extraction from various observables. The
determination from $\tau$ decays is  important, since it provides an accurate
extraction at low energies, close to the limit of validity of perturbative QCD.

The general framework for the determination of $\alpha_s$ from the ratio
\begin{equation}
\label{eq:RtauEq1}
R_\tau \,=\, \frac{\Gamma \left[ \tau^- \to \nu_\tau {\rm hadrons}(\gamma)
\right]}{\Gamma \left[ \tau^- \to \nu_\tau e^-\bar \nu_e(\gamma)  \right]}
\,=\, 3.6280 \pm 0.0094\,\,\, \mbox{\cite{HFAG12}}
\end{equation}
was developed about 20 years ago~\cite{BNP92}. Theoretically, $R_\tau$ can be
expressed as a weighted integral of the measured hadronic spectral functions
that runs over the hadronic invariant mass squared $s$ of the hadronic final
state from threshold up to $m_\tau^2$. The relevant weight function,
$w_\tau(x)$, is obtained from the kinematics of the decay. However, the use
of QCD at very low energies is impractical. Therefore, one resorts to a
finite-energy sum rule (FESR) where the theoretical counter-part of $R_\tau$
is evaluated as a contour integral in the complex-energy plane with
$|s|=m_\tau^2$. A particularity of this observable is that non-perturbative
effects, although small, cannot be neglected. The perturbative QCD result
must be supplemented with power corrections organised in an operator product
expansion (OPE). With the data from the ALEPH~\cite{ALEPH,ALEPH2,Davier} and
OPAL~\cite{OPAL} collaborations, as well as progress on the theory side, the
precision on $\alpha_s(m_\tau)$ is impressive: the advocated uncertainties
that two decades ago were around~$11\%$~\cite{BNP92}, are now of the order of
$2.5\%$~\cite{Davier} for the most optimistic analysis. However, the results
obtained by different groups are sometimes barely compatible, which suggests
that the details of the different analyses need to be scrutinised.

In the theoretical description two ingredients are needed. The first of them
is the perturbative QCD contribution, the second are the non-perturbative
effects. At present, two theoretical obstacles obstruct progress on the theory
side. First, the renormalisation group improvement of the perturbative series
remains controversial --- we return to this subject below. Second, the treatment
of non-perturbative effects (encoded in the OPE) in some of the existent
analyses was shown to be inconsistent~\cite{MY08}. A possible solution to this
problem, proposed in~\cite{MY08, M7C1, M7C2}, is the inclusion of the so-called
duality violations (DVs) in the analysis framework. These are related to the
fact that the OPE fails to describe the spectral functions near the Minkowski
axis, where resonance effects may become important and local quark-hadron
duality is violated. In the past, the standard assumption was that DVs could
be disregarded due to the kinematical suppression of contributions from this
problematic region. Progress in modelling
DVs~\cite{DVsRussians,DVsRussians2,DVsRussians3,CGP05,CGP08,CGP,MJ11} made it possible
to include them in the $\alpha_s$ analyses without reliance on external
input~\cite{M7C1,M7C2}, and to test the above assumption.

In the present work we focus on the perturbative contribution. One of the
main sources of uncertainty in the theory of hadronic $\tau$ decays is the
renormalisation group (RG) improvement of the perturbative series. The most
widely employed prescriptions are fixed-order perturbation theory (FOPT, see
for instance ref.~\cite{Jamin05,BJ2008}) and contour-improved perturbation
theory (CIPT)~\cite{CIPT,PLD92}. Employing these prescriptions at a finite
order in perturbation theory leads to differing values for $\alpha_s$. The
inclusion of the
recently computed $\alpha_s^4$ correction~\cite{bck08} to $R_\tau$ rendered
the discrepancy between FOPT and CIPT even more pronounced. Since then,
several works have dealt with the RG improvement of the
series~\cite{BJ2008,SM,CF,DGM,Cvetic:2010ut,AAC}. A difficulty common to
all these works is that conclusions in favour of FOPT or CIPT (or a third
prescription) depend on implicit or explicit assumptions on the yet unknown
higher order coefficients of the Adler function. In particular, the aim of
ref.~\cite{BJ2008} was to construct a plausible model for the perturbative
series in higher orders incorporating only general features of the leading
renormalon singularities of the Borel-transformed Adler function. This should
be sufficient to describe the perturbative coefficients at intermediate and
high orders, augmented by some polynomial terms to take care of the first few
coefficients which are not yet dominated by (pre-)asymptotic behaviour. After
matching of the model to the known coefficients of the Adler function in QCD,
the main conclusion of ref.~\cite{BJ2008} was that FOPT is to be preferred over
CIPT, since at order $\alpha_s^4$ and in the region of its smallest terms, FOPT
provides a closer approach to the resummed series than CIPT. 

How general is this conclusion? A short-coming of ref.~\cite{BJ2008} and other
recent works (with the exception of ref.~\cite{DGM}) is that the analysis
was done solely for the kinematical weight $w_\tau$. This is not entirely
satisfactory because the $\alpha_s$ determinations employ --- and often require
--- several different weight functions, if only to extract $\alpha_s$ together
with the non-perturbative condensates and DV parameters from the same
self-consistent analysis. In fact, any analytic weight function $w_i(x)$ gives
rise to a valid FESR, and different $w_i(x)$ emphasise different energy regions
of the experimental data and different contributions of the theoretical
description. In the analyses of $\alpha_s$ found in
the literature, several different moments have been used. The enhancement or
the suppression of condensates and DV contributions have been the guiding
principle in choosing these moments. Still, little attention has been devoted
to the moment dependence of the convergence properties of the perturbative
series and we aim to fill this gap here.
In the present paper we therefore pursue the FOPT/CIPT comparison 
using the methods of ref.~\cite{BJ2008}, and ask whether the preference for
one or the other depends on specific features of the weight function, or the
assumptions on the Adler function coefficients; whether the kinematic weight
is special, or all moments are alike.

The outline is as follows. After setting up the necessary notation, we study
how the convergence properties of the perturbative expansion depend on the
choice of the weight function that defines the moment. We try, as much as
possible, to remain model independent by making use not only of the reference
model of ref.~\cite{BJ2008}, but also employing an extreme case where CIPT is,
by construction, preferred over FOPT for the kinematical weight function. We 
show that with respect to convergence properties and the FOPT/CIPT comparison
the moments can be divided into several classes, whose global features are
manifestations of simple properties of the weight functions and assumptions on
the Adler function series.  We conclude that certain weight functions should
be more suitable for $\alpha_s$ analyses than others. We then study the
robustness of the model proposed in ref.~\cite{BJ2008} in the light of the
criticism presented in ref.~\cite{DGM} and provide further plausibility
arguments in favour of the adopted procedure. Finally, in the last section we study
the consistency between the moments by performing simplified $\alpha_s$
determinations from single-moment fits.

\section{Theoretical framework}
\label{TheoreticalFramework}

The total decay rate of the $\tau$ lepton into hadrons, eq.~(\ref{eq:RtauEq1}),
can be separated experimentally into three components: the vector, $R_{\tau,V}$,
and axial-vector, $R_{\tau,A}$, arising from the decays into light quarks
through the $(\bar u d)$-quark current, and contributions with net strangeness,
$R_{\tau,S}$, from the ($\bar u  s$)-quark current. Hence
\begin{equation}
R_\tau \,=\, R_{\tau,V} + R_{\tau,A} + R_{\tau,S} \,.
\end{equation}
In determinations of $\alpha_s$, the focus is on the non-strange contributions,
because power corrections are largest in the strange sector, while they are
suppressed by the light $u$- and $d$-quark masses for $R_{\tau,V}$ and
$R_{\tau,A}$. For this reason, in the following we restrict ourselves to the
two latter channels.

The ratios $R_{\tau,{V/A}}$ can be expressed in terms of integrals of the
spectral functions $\Im\,\Pi^{(1)}_{V/A}$ and $\Im\,\Pi^{(0)}_{V/A}$ as
\begin{equation}
\label{Rtauth}
R_{\tau,{V/A}} \,=\, 12\pi S_{\rm EW} |V_{ud}|^2 \!\int\limits_0^{m_\tau^2}
\frac{ds}{m_\tau^2}\,\biggl( 1-\frac{s}{m_\tau^2}\biggr)^{\!2}
\biggl[\biggl(1+2\frac{s}{m_\tau^2}\biggr)
\Im\,\Pi^{(1)}_{V/A}(s)+\Im\,\Pi^{(0)}_{V/A} (s) \,\biggr] \,.
\end{equation}
In the last equation, $S_{\rm EW}$ is an electroweak
correction~\cite{MS88,BL90,Erl04}, and $V_{ud}$ is the quark-mixing matrix
element~\cite{Vud}. Theoretically, the relevant two-point functions whose
spectral functions enter eq.~(\ref{Rtauth}) are
\begin{equation}
\label{PiVAmunu}
\Pi_{V/A}^{\mu\nu}(p) \,\equiv\,  i\!\int \! dx \, e^{ipx} \,
\langle\Omega|\,T\{ J_{V/A}^{\mu}(x)\,J_{V/A}^{\nu}(0)^\dagger\}|
\Omega\rangle\,,
\end{equation}
with $\ket{\Omega}$ being the physical vacuum, and the $V$ and $A$ currents
are $J_{V/A}^{\mu}(x)= (\bar u\gamma^\mu(\gamma_5) d)(x)$. These correlators
assume the standard decomposition into transversal and longitudinal components 
which was employed in writing eq.~(\ref{Rtauth}).

One then makes use of the fact that the exact correlation functions are
analytic in the complex $s$-plane except for a cut along the real axis. This
property allows one to write eq.~(\ref{Rtauth}) as a counter-clockwise contour
integral along the circle $|s|=s_0$
\begin{equation}
\label{Rtaucon}
R^{w_i}_{V/A}(s_0) \,=\, 6\pi i\,S_{\rm EW} |V_{ud}|^2 \!\!\!
\oint\limits_{|s|=s_0} \frac{d s}{s_0}\, w_i\left(s\right) \biggl[\,\Pi^{(1+0)}_{V/A}(s) +
\frac{2s}{(s_0+2s)}\,\Pi^{(0)}_{V/A}(s) \,\biggr] \,.
\end{equation}
In writing the last equation we have performed two generalisations. First, we
are using a generalised analytic weight function $w_i(s)$, second, the integral
is performed up to an arbitrary energy $s_0\leq m_\tau^2$. In the notation of
eq.~(\ref{Rtaucon}), the particular case of eq.~(\ref{Rtauth}) corresponds to
$R^{w_\tau}_{V/A}(m_\tau^2)$ with $s_0=m_\tau^2$ and
\begin{equation}
\label{wtau}
w_\tau(s) \,=\, \Big(1-\frac{s}{m_\tau^2}\Big)^{2}\Big(1+2\frac{s}
{m_\tau^2}\Big) \,.
\end{equation}

For large enough $s$, the contributions to $\Pi^{(J)}(s)$ can be organised in
an operator product expansion: a series of local gauge-invariant operators of
increasing dimensions times the appropriate inverse powers of $s$. In this
framework, the purely perturbative part in the chiral limit can be associated
with the dimension-zero operator, whereas dimension-2 contributions arise from
the quark mass corrections.\footnote{In the case at hand, namely $u$ and $d$
quarks only, the dimension-2 corrections are tiny.} The first non-trivial
operators arise at dimension 4, namely, the quark and gluon condensates. The
OPE is expected to be well behaved along the contour $|s|=s_0$ (for $s_0$
sufficiently large) except close to the positive real axis.  Therefore, in the
general case, $R^{w_i}_{V/A}(s_0)$ obtains a contribution from corrections due
to the break-down of the OPE close to real $s>0$. This term is the
aforementioned DV contribution. Weight functions $w_i(s)$ that contain
one or more zeros at $s=s_0$, such as the kinematical $w_\tau$ in
eq.~(\ref{wtau}), tend to suppress the contribution of DVs.

The different components of $R_{V/A}^{w_i}$ can be collected in the following
expression
\begin{equation}
\label{RtauDeltas}
R_{V/A}^{w_i}(s_0) \,=\, \frac{N_c}{2}\,S_{EW} |V_{ud}|^2 \biggl[\,
\delta^{\rm tree}_{w_i} + \delta^{(0)}_{w_i}(s_0) + 
\sum_{D\geq 2}\delta^{(D)}_{w_i,V/A}(s_0) +\delta^{\rm DV}_{w_i,V/A}(s_0)
\,\biggr] \,.
\end{equation}
In the last equation, $\delta^{\rm tree}_{w_i}$ and $\delta^{(0)}_{w_i}$ are
the perturbative terms,\footnote{Henceforth, we omit the $s_0$ dependence in
the terms of the r.h.s of eq.~(\ref{RtauDeltas}).} of which $\delta^{(0)}_{w_i}$
contains the $\alpha_s$ corrections. Since $\delta^{\rm tree}_{w_i}$ and
$\delta^{(0)}_{w_i}$ do not depend on the flavour, in the chiral limit they
are the same for vector and axial-vector correlators, and correspond to the
perturbative series for the correlator $\Pi^{(1+0)}_{V/A}(s)$. The
contributions from the quark masses, as well as that of the operators with
$D>2$, are encoded in the terms $\delta^{(D)}_{w_i,V/A}$, while the DV
contributions are represented by $\delta^{\rm DV}_{w_i,V/A}$. In this work
we are interested in the convergence properties of the purely perturbative
corrections, and thus our focus is on~$\delta^{(0)}_{w_i}$.

The correlator $\Pi^{(1+0)}$ is not RG invariant and contains scale- and
scheme-dependent contributions. However, the Cauchy integral in
eq.~(\ref{Rtaucon}) is insensitive to all $s$-independent terms in the
correlators. Without loss of generality, one can work with a renormalisation
invariant quantity, known as the Adler function, and defined through
\begin{equation}
\label{D10D0}
D^{(1+0)}(s) \,\equiv\, -\,s\,\frac{d}{ds}\,\Pi^{(1+0)}(s) \,.
\end{equation}
Using partial integration, and performing the substitution $x=s/s_0$, one can
write
\begin{equation}
\label{delta0Int}
\delta^{(0)}_{w_i}  \,=\, -\,2\pi i\!\!\! \oint\limits_{|x|=1}\!\!
\frac{d x}{x}\, W_i(x)\,D_{\rm pert}^{(1+0)}(s_0 x) \,,
\end{equation}
where ``pert'' denotes the perturbative part of the Adler function in the
chiral limit and the weight function $W_i(x)$ is obtained from $w_i(x)$ by
the integral $W_i(x) = 2\int_x^1 dz\, w_i(z)$.

In full generality, the perturbative Adler function admits the following
expansion:
\begin{equation}
\label{Ds}
D^{(1+0)}_{\rm pert}(s) \,=\, \frac{N_c}{12\pi^2} \sum\limits_{n=0}^\infty
a_\mu^n \sum\limits_{k=1}^{n+1} k\, c_{n,k}\,L^{k-1} \,,
\qquad L\equiv \log \frac{-s}{\mu^2} \,,
\end{equation}
with $a_\mu \equiv a(\mu^2)\equiv \alpha_s(\mu)/\pi $, $\mu$ is the
renormalisation scale, and $N_c$ the number of colours. Imposing the RG
invariance of the above equation, one may consider as independent only the
coefficients $c_{n,1}$. The other coefficients $c_{n,k}$, with $k=2,3,...,n+1$,
can be obtained in terms of the $c_{n,1}$ and $\beta$-function
coefficients.\footnote{We follow the convention of ref.~\cite{BJ2008},
 i.e. $\beta(a_\mu)\equiv \mu da_\mu/d\mu = \sum_{k=1}\beta_ka_\mu^{k+1}$.
The first coefficient is then $\beta_1 = 11N_c/6 - N_f/3$.} (Explicit
expressions for some of the coefficients $c_{n,k}$ can be found in eq.~(2.11)
of ref.~\cite{BJ2008}.) At $N_c=N_f=3$ the numerical values of the known
coefficients $c_{n,1}$ are
\begin{equation}
c_{0,1} \,=\, c_{1,1} \,=\, 1 \,, \quad
c_{2,1} \,=\, 1.640 \,, \quad
c_{3,1} \,=\, 6.371 \,\mbox{\cite{gkl91,ss91}}\,, \quad
c_{4,1} \,=\, 49.076 \,\,\mbox{\cite{bck08}}\,.
\label{exactcoeffs}
\end{equation}
Fully analytic results for the coefficients can be found in ref.~\cite{bck08}.
Based on a geometrical growth of the terms in the perturbative expansion of
$\delta^{(0)}_{w_\tau}$, in ref.~\cite{BJ2008} the estimate 
\begin{equation}
\label{c51}
c_{5,1} \,\approx\, 283
\end{equation}
was put forward for the next term in the series.  This estimate was
corroborated by the model introduced in ref.~\cite{BJ2008} and we will also
employ it in our work.

Inserting the general expansion of the Adler function, eq.~(\ref{Ds}), into
the expression for $\delta^{(0)}_{w_i}$, eq.~(\ref{delta0Int}), yields 
\begin{equation}
\label{del0}
\delta^{(0)}_{w_i} \,=\, \sum\limits_{n=1}^\infty  \sum\limits_{k=1}^{n}
k\,c_{n,k} \;\frac{1}{2\pi i}\!\!\oint\limits_{|x|=1} \!\! \frac{dx}{x}\,
W_i(x) \log^{k-1}\biggl(\frac{-s_0 x}{\mu^2}\biggr)a_\mu^n \,.
\end{equation}
Since the Adler function satisfies a homogeneous RG equation, the above
expression for $\delta^{(0)}_{w_i} $ is $\mu$-independent. One has the freedom
of setting the scale $\mu$ in a convenient way.

The fixed-order prescription corresponds to $\mu^2=s_0$. In this case,
the coupling is calculated at a fixed scale and can be taken outside the
integral. However, the logarithms remain to be integrated along the contour.
The result can be cast into
\begin{equation}
\label{del0FO}
\delta^{(0)}_{{\rm FO},w_i} \,=\, \sum\limits_{n=1}^\infty a(s_0)^n
\sum\limits_{k=1}^{n} k\,c_{n,k}\,J_{k-1}^{{\rm FO},w_i} \,,
\end{equation}
where the integrals are given by
\begin{equation}
\label{Jl}
J_n^{{\rm FO},w_i} \,\equiv\, \frac{1}{2\pi i} \!\!\oint\limits_{|x|=1} \!\!
\frac{dx}{x}\,W_i(x) \log^n(-x) \,.
\end{equation}
For polynomial moments, these integrals can be performed analytically. Explicit
expressions for the particular case of the kinematic weight function can be
found in ref.~\cite{BJ2008}.

In contour-improved perturbation theory~\cite{CIPT,PLD92}, 
the logarithms that remain in the
FO prescription are summed with the choice $\mu^2=-s_0 x$ before calculating
the contour integral. This procedure implies that the contour integrals have
to be performed over the running $\alpha_s$ in the complex plane
\begin{equation}
\label{del0CI}
\delta^{(0)}_{{\rm CI},w_i} \,=\, \sum\limits_{n=1}^\infty c_{n,1}\,
J_n^{{\rm CI},w_i}(s_0) \,,
\end{equation}
where the integrals, that can only be computed numerically, are given by
\begin{equation}
\label{Jna}
J_n^{{\rm CI},w_i}(s_0) \,\equiv\, \frac{1}{2\pi i} \!\!\oint\limits_{|x|=1}
\!\! \frac{dx}{x}\,W_i(x)\,a^n(-s_0 x) \,.
\end{equation}
CIPT resums the running of the QCD coupling along the contour of integration.
Consequently, at each order $n$, only the coefficient $c_{n,1}$ enters the
expression.

\section{Models for the Adler function}
\label{RenormalonModels}

In order to discuss the behaviour of the perturbative expansion of the spectral
moments and to compare FO to CI perturbation theory, we need an ansatz for the
coefficients $c_{n,1}$ of the Adler function, which is the dynamical input
common to all moments, beyond $n=4$. In this section we introduce the models
for the series that we use later on. A caveat needs to be spelled out at this 
point. Going beyond the exactly known coefficients $c_{n,1}$ requires
assumptions, usually based on some form of regularity of the series, which
might simply be wrong. The series might have outliers at some order, and we
will never know. The best we can do is to state the assumptions clearly, to
provide supporting arguments where they exist, and to explore the consequences.
Two diverging assumptions may be distinguished:
\begin{itemize}
\item The high-order coefficients $c_{n,1}$ beyond $n=4$ are not important,
and can be neglected. The RG improvement still generates a non-trivial series
expansion of the spectral moments to all orders through the dependence of
$c_{n,k}$ with $k>1$ on the known $c_{n,1}$. In a sense this is the assumption
underlying CIPT, which assumes that the running coupling terms are dominant
and therefore should be summed.
\item The high-order coefficients $c_{n,1}$ beyond $n=4$ are essential, since
they diverge factorially for large $n$, thus overcoming the geometric growth
of the running coupling terms. Since some knowledge exists on the general
structure of this divergence (reviewed in ref.~\cite{Renormalons}), this
information  can and should be included.
\end{itemize}
The two main models that we discuss in this section can be viewed as
representatives of these assumptions. In addition we also briefly review the
result in the large-$\beta_0$ approximation,\footnote{For historical reasons,
we speak about the ``large-$\beta_0$'' approximation, although in the notation
employed in this work, the leading coefficient of the $\beta$-function is
termed $\beta_1$.} which, since it is based on a well-defined formal limit
($N_f\to -\infty$) of QCD, provides a useful toy model to which we shall return
in section~\ref{Plausibility}. In the context of tau decays, this toy model
has been studied in refs.~\cite{bbb95,neu95}.

We start by giving a number of definitions and establishing the notation. We
define a new function $\wh D(s)$ related to the Adler function by
\begin{equation}
\label{Rs}
\frac{12\pi^2}{N_c}\,D^{(1+0)}_V(s) \,\equiv\, 1 + \wh D(s) \,\equiv\, 1 +
\sum\limits_{n=0}^\infty r_n \,\as(\sqrt{s})^{n+1} \,.
\end{equation}
The coefficients $c_{n,1}$ of $D^{(1+0)}_V$ are related to those of $\wh D(s)$
by $c_{n,1}=\pi^ n r_{n-1}$. The Borel transform of the above series is
defined by
\begin{equation}
\label{BRt}
B[\wh D](t) \,\equiv\, \sum\limits_{n=0}^\infty r_n\,\frac{t^n}{n!} \,.
\end{equation}
One can then define the Borel integral of the series as ($\alpha$ positive)
\begin{equation}
\label{Dalpha}
\wh D(\alpha) \,\equiv\, \int\limits_0^\infty dt\,{\rm e}^{-t/\alpha}\,
B[\wh D](t)\,,
\end{equation}
which has the same series expansion in $\alpha$ as $\wh D(s)$ has in
$\as(\sqrt{s})$. The last integral, $\wh D(\alpha)$, if it exists, gives the
Borel sum of the original divergent series eq.~(\ref{Rs}). An important point
in the case of the Adler function is that $B[\wh D](t)$ contains singularities
on the positive real axis which forces one to adopt a procedure to define the
integral~$\wh D(\alpha)$. The choice of the procedure introduces an ambiguity.
We discuss this point in more detail below.

\subsection[Large-$\beta_0$ model]{\boldmath Large-$\beta_0$ model}

In the context of the large-$\beta_0$ approximation, it has been shown by
resumming bubble-chain diagrams that the Borel-transformed Adler function has
infrared (IR) and ultraviolet (UV) renormalon poles at positive and negative
integer values of the variable $u=\beta_1 t/(2\pi)$,
respectively~\cite{ben93,bro93}. (Except at the value $u=1$.) The IR renormalon
poles are related to the power corrections in the OPE, while the leading UV
renormalon dictates the large-order behaviour of the series. (For a review see
ref.~\cite{Renormalons}).

The main result of refs.~\cite{ben93,bro93} is that in the large-$\beta_0$
approximation the Borel transformed Adler function can be written
as~\cite{bro93}
\begin{equation}
\label{BRuk}
B[\widehat D](u) \,=\, \frac{32}{3\pi}\,
\frac{{\rm e}^{-C u}}{(2-u)}\,
\sum\limits_{k=2}^\infty\,\frac{(-1)^k k}{[k^2-(1-u)^2]^2} \,,
\end{equation}
where the constant $C$ is scheme dependent and cancels the scheme dependence
of $\alpha_s$ in eq.~(\ref{Dalpha}), such that $\widehat D(s)$ is scheme
independent. (In the $\MSb$-scheme $C=-5/3$.)  In large-$\beta_0$, all the UV
renormalon poles at $u=-1,-2,...$, are double poles, as are all the IR poles
at $u=3,4,...$. The only exception is the IR pole at $u=2$, which is simple.
This follows from the fact that the operator $\alpha_s GG$ has no anomalous
dimension in the large-$N_f$ limit. The absence of an IR renormalon pole at
$u=1$ stems from the fact that no dimension-2 operator contributes to the OPE.
The coefficients $c_{n,1}$ in this case can be obtained from eq.~(\ref{BRuk})
by expanding in $u$ and performing the Borel integral term by term. The first
12 coefficients can be found in table~1 of ref.~\cite{BJ2008}. An interesting
feature of the large-$\beta_0$ result, in apparent coincidence with the full
QCD series (\ref{exactcoeffs}),  is that the asymptotically dominant
sign-alternation from the UV pole at $u=-1$ is delayed in the conventionally
adopted $\MSb$-scheme. In intermediate orders the series coefficients are
governed by the fixed-sign contributions from the $u=2$ pole, whose residue is
a factor $e^{-3 C}$ larger.

\subsection{Reference model}

In full QCD we do not have the equivalent of eq.~(\ref{BRuk}). On the other
hand, the structure of the OPE and general RG arguments allow one to determine
the position and strength of the singularities, which evolve from poles into
branch cuts~\cite{BJ2008,mue85,Beneke:1993ee}, though not their residues. In
general, the IR and UV singularities are described by the following structures
\begin{eqnarray}
\label{BR3PIR}
B[\wh D_p^{\rm IR}](u) \,&\equiv&\, 
\frac{d_p^{\rm IR}}{(p-u)^{1+\tilde\gamma}}\,
\Big[\, 1 + \tilde b_1 (p-u) + \tilde b_2 (p-u)^2  + \cdots\,\Big] \,, \nn \\
B[\wh D_p^{\rm UV}](u) \,&\equiv&\, \frac{d_p^{\rm UV}}{(p+u)^{1+\bar\gamma}}\,
\Big[\, 1 + \bar b_1 (p+u) + \bar b_2 (p+u)^2 + \cdots\,\Big] \,,
\end{eqnarray}
where the constants $\tilde \gamma$, $\tilde b_i$, $\bar\gamma$, and $\bar b_i$
of a pole at $p$ depend on anomalous dimensions of operators in the OPE as
well as $\beta$-function coefficients (the explicit expressions are given in
section~5 of ref.~\cite{BJ2008}). When performing the integral~(\ref{Dalpha})
one needs to circumvent the IR singularities along the real axis. A prescription
to define the integral is needed which introduces an ambiguity in the Borel
resummed result. The ambiguity is expected to be cancelled by exponentially
small terms in $\alpha_s$ or, due to the running of the coupling, by power
corrections.  The treatment of the Borel integral in the presence of these
singularities is discussed in appendix~A of~\cite{BJ2008}.

The model for the Adler function constructed in ref.~\cite{BJ2008} (henceforth
called \emph{reference model}, or simply RM) is based on the assumption that it
makes sense to merge the exactly known low-order behaviour to the leading and
sub-leading asymptotics generated by singularities of the Borel transform.
Since the known coefficients do not display the asymptotic sign alternating
pattern, the leading, first UV singularity should be sufficient. On the other
hand, if the intermediate orders are governed by fixed-sign behaviour, at least
the first two IR renormalon singularities should be included in the model.
Based on these considerations, the ansatz reads
\begin{equation}
B[\widehat D](u) \,=\, B[\widehat D_1^{\rm UV}](u)+ B[\widehat D_2^{\rm IR}](u)
+ B[\widehat D_3^{\rm IR}](u)+ d_0^{\rm PO} + d_1^{\rm PO}\,u\label{CMBJ} \,,
\end{equation}
where the renormalon singularities are described by the formulae of
eq.~(\ref{BR3PIR}). 

This ansatz is then matched to the known coefficients of the Adler function in
QCD as follows. First, the known coefficients $c_{3,1}$, $c_{4,1}$, and the
estimated coefficient $c_{5,1}$, are used to fix the residua of the three
renormalon singularities. The polynomial terms are then fixed in order to
reproduce the lowest order coefficients $c_{1,1}$ and $c_{1,2}$. The resulting
parameters are given in eq.~(6.2) of~\cite{BJ2008} and the first line of
table~\ref{CMvsDGM} in section~\ref{Plausibility}, and take ``reasonable''
values. One is then in a position to perform the Borel integration in order
to ascribe a resummed value to the asymptotic Adler function series. The
higher-order coefficients $c_{n,1}$ can be derived and the behaviour of FOPT
and CIPT series can be compared to the resummed one.

The main conclusion of ref.~\cite{BJ2008} is that under the above assumptions 
FOPT is clearly preferred over CIPT for $w_\tau$. The CIPT series displays a
faster convergence but fails to give a good approximation to the Borel resummed
result in the sense of an asymptotic series. From FOPT, on the other hand, it
is possible to extract a good approximation to the Borel resummed value in spite
of the slower convergence of the series. The reason for this observation can be
traced back to cancellations that are missed by the CIPT series. To understand
this we rewrite $\delta^{(0)}_{{\rm FO},w_i}$ of eq.~(\ref{del0FO}) as
\begin{equation}
\label{delFOcg}
\delta^{(0)}_{{\rm FO},w_i} \,=\, \sum\limits_{n=1}^\infty 
\left[c_{n,1}\delta_{w_i}^{\rm tree} + g_n^{[w_i]}\right] a(s_0)^n \,,
\end{equation}
with 
\begin{equation}
g_n^{[w_i]} \,=\, \sum_{k=2}^n k \,c_{n,k} J_{k-1}^{{\rm FO},w_i} \,.
\end{equation}
The $c_{n,1}$ series is simply the Adler function series multiplied by
$\delta_{w_i}^{\rm tree}$, while the contour integration of the running
coupling effects is fully contained in the $g_n^{[w_i]}$ series. In
eq.~(\ref{delFOcg}) the tree-level contribution arises because
$J_0^{{\rm FO},w_i}=W_i(0)=\delta_{w_i}^{\rm tree}$. FOPT treats the $c_{n,1}$
and the $g_n^{[w_i]}$ series on an equal footing. Comparing this decomposition
with the CIPT result, eq.~(\ref{del0CI}), one observes that in CIPT the
$g_n^{[w_i]}$ series is resummed to all orders while the $c_{n,1}$ series is
used only up to a finite order $n$. An important model-independent feature 
of the QCD series, which follows from the OPE and the form of the moment weight
function, is that there are large cancellations of $n!$ divergences between the
$c_{n,1}$ and $g_n^{[w_i]}$ series. These cancellations are particularly strong
when the series is dominated by the $u=2$ singularity, and for the kinematic 
weight.\footnote{In large-$\beta_0$ this is shown analytically for $w_\tau$ in
ref.~\cite{BJ2008}. An important point discussed in the next section is the
moment dependence of the cancellations.} In such a scenario, it is mandatory
to combine  $c_{n,1}$ and $g_n^{[w_i]}$ order by order in $n$, lest the
cancellations do not take place. Since CIPT treats the orders incoherently, it
misses the cancellations and runs into the sign alternating asymptotic regime
earlier. In FOPT, on the other hand, the cancellations suppress the divergence
and allow FOPT to approach the Borel result. In table~\ref{cancellations}, we
show as an example the cancellations in the case of the kinematical moment
$w_\tau$. However, note that they are not imposed in the RM. Rather, the
matching procedure to the QCD series gives the expected weight to the leading 
IR pole. If the residue of the IR pole $u=2$ had turned out to be tiny, this
would have made the cancellations almost non-existent.

\begin{table}[t]
\begin{center}{\small
\begin{tabular}{ r l l c | l l c }
\toprule
&\multicolumn{3}{c}{Reference model} & \multicolumn{3}{c}{Alternative model} \\
\midrule
$n$ & $\quad c_n$ & $\quad g_n^{[w_\tau]}$ &
$\left(c_n+g_n^{[w_\tau]}\right)/c_n$ & $\quad c_n$ & $\quad g_n^{[w_\tau]}$ &
$\left(c_n+g_n^{[w_\tau]}\right)/c_n$ \\
 \midrule
 4 & $49.1$ & $\phantom{-}78.0$ & $2.59$ & $49.1$ & $\phantom{-}78.0$ &
 $\phantom{-}2.59$ \\
 5 & $283$ & $\phantom{-}307.8$ & $2.09$ & $283$ & $\phantom{-}307.8$ &
 $\phantom{-}2.09$ \\
 6 & $3275.4$ & $-807.3$ & $0.75$  & $2148.3$ & $-807.3$ & $\phantom{-}0.62$ \\
 7 & $18,758$ & $-10,398$ & $0.45$ & $11,801$ & $-34,489$ & $-1.92$ \\
 8 & $388,442$ & $-329,054$ & $0.15$ & $150,508$ & $-592,196$ & $-2.93$ \\
 9 & $919,121$ & $-232,718$ & $0.75$ & $215,264$ & $-5.1\times 10^6$ & $-22.8$ \\
10 & $8.4\times 10^7$ & $-7.3\times 10^7$ & 0.12 &  $2.4\times 10^7$ &
 $-6.4\times 10^7$ & $-1.69$ \\
 \bottomrule
\end{tabular}
\caption{Cancellations between the $c_n$ and the $g_n^{[w_\tau]}$ series for
orders $4\leq n\leq 10$ for the kinematical moment, $w_\tau$, in the reference
model and in the alternative model, eqs.~(\ref{CMBJ}) and~(\ref{ArtifModel}).
\label{cancellations}}
}\end{center}
\end{table}

\subsection{Alternative model}

To make this feature clearer, we can {\it artificially} suppress the leading
IR pole. Let us consider a model for the Borel transformed Adler function
where the IR singularity at $u=2$ is removed and another at $u=4$ is added:
\begin{equation}
\label{ArtifModel}
B[\widehat D](u) \,=\, B[\widehat D_1^{\rm UV}](u)+ B[\widehat D_3^{\rm IR}](u)
+B[\widehat D_4^{\rm IR}](u) + d_0^{\rm PO} + d_1^{\rm PO}\,u  \,.
\end{equation}
In this model, the aforementioned cancellations do not take place by
construction.  We refer to this model as the \emph{alternative model} (AM). An
analogous matching procedure can be carried out yielding the following values
for the parameters:
\begin{equation}
\begin{array}{lll}
d_{3}^{\rm IR} \,=\, 66.18 \,, &\qquad d_4^{\rm IR} \,=\, -289.71 \,, &\qquad
d_{1}^{\rm UV} \,=\, -5.21\times 10^{-3} \,,\\
d_0^{\rm PO} \,=\, 2.15 \,, &\qquad d_1^{\rm PO} \,=\, 4.01 \times 10^{-1} \,.
& \\
\end{array}
\end{equation}
Here, as we show in the sequel, CIPT is able to approach the Borel
resummed result while FOPT exhibits oscillations around this value.
Table~\ref{cancellations} shows that the cancellations between the $c_n$ and
$g_n^{[w_i]}$ series no longer take place in this model. The table also shows
a slower growth of the $c_{n,1}$ in this model, and a dominance of the $g_n$
terms up to $n=10$, which therefore realises a situation where running
coupling effects are dominant. We use this model as an example where CIPT is,
by construction, superior to FOPT, at least for the kinematical weight. This
provides a way to assess a possible model dependence in our conclusions.
Nevertheless, we emphasise that we find it unlikely that the Adler function in
QCD behaves as the AM, since there is no known mechanism that would naturally
suppress the $u=2$ singularity.

\section{Moment analysis}
\label{Moments}

The determination of $\alpha_s$ and condensates from the analysis of $\tau$
hadronic spectral functions is based on sum rules obtained by equating
eqs.~(\ref{Rtauth}) and (\ref{Rtaucon}). In the former, to perform the integral
along the real axis, the experimental spectral functions are used. In
eq.~(\ref{Rtaucon}), the theoretical description of the correlators is employed
in the contour integration. An important aspect of these sum rules is that one
still has the freedom of choosing any analytic weight function $w_i(x)$, as
well as any point $s_0\leq m_\tau^2$ (as long as $s_0$ is large enough for the
OPE and the perturbative expansion to make sense). On the experimental side, it
is obvious that a given weight function enhances the regions of the spectrum
where it has peaks. On the theory side, the relative contributions of the
different $\delta$'s in eq.~(\ref{RtauDeltas}) are strongly dependent on the
choice of $w_i(x)$. For example, as already mentioned, moments of functions
$w_i$ with zeros at $s=s_0$ suppress $\delta^{\rm DV}_{w_i}$. These are known
in the literature under the name {\it pinched moments}. A monomial term of
the type $x^k$ in $w_i$, on the other hand, implies the monomial $x^{k+1}$ in
$W_i$ defined after eq.~(\ref{delta0Int}), which enhances (or, rather, does not
suppress) the contribution of the condensate of dimension $D=2(k+1)$ as well as
the factorial divergence from the IR renormalon singularity at $u=k+1$. Let us
analyse, as an illustrative example, the kinematical moment
\begin{equation}
w_\tau \,=\, (1-x)^2 (1+2x) = 1-3x^2+2x^3.
\end{equation}
It receives its larger contributions from $\delta^{\rm tree}_{w_\tau}$,
$\delta^{(0)}_{w_\tau}$, $\delta^{(6)}_{w_\tau}$, and $\delta^{(8)}_{w_\tau}$.
The first two arise mainly from the 1 in $w_\tau$, whereas
$\delta^{(6)}_{w_\tau}$ and $\delta^{(8)}_{w_\tau}$ arise from the terms
$-3x^2$ and $2x^3$, respectively. The double zero at $x=1$ suppresses 
$\delta^{\rm DV}_{w_\tau}$, while the absence of other monomial terms suppresses
the condensates with $D=4$ as well as with $D\geq10$. (The mass corrections,
$\delta^{(2)}_{w_\tau}$, are negligible due to the smallness of the quark
masses.)

In order to extract $\alpha_s$, a number of condensates, and the DV parameters
from the data sets one needs more than one observable. It has become standard
to use a set of several weight functions $w_i$ in order to perform a combined
fit to their --- not statistically independent --- moments. In table~\ref{tab:ws}
we collect the weight functions investigated in this work. Most of them have
been employed in at least one of the recent analyses of hadronic $\tau$ spectral
functions. In this table, the first five rows are the building blocks for the
other polynomial weight functions. The second set are pinched weight-functions
that contain a $1$ followed by powers of $x$. The third block contains pinched
weight-functions that do not have the $1$ and start directly with some power of
$x$. The idea behind the moments that were used in the existent analyses was
mainly the enhancement or the suppression of condensates and DV contributions.
In combined fits to sets of moments (e.g. refs.~\cite{Davier,OPAL,MY08,M7C1,M7C2}),
the final value of $\alpha_s$ receives contributions from the perturbative
terms of all moments employed. Therefore, in order to achieve a trustworthy
determination of the coupling, it is desirable to understand the convergence
properties of the perturbative component for all the moments employed in the
$\alpha_s$ analysis.

\begin{table}[t]
\begin{center}
\begin{tabular}{l l l l}
\toprule
 $i$ & $w_i(x)$ & $\delta^{\rm tree}_{w_i}$ & refs. \\
 \midrule
  1 & 1     & 2 & \cite{M7C1, M7C2}\\
  2 & $x$   & 1 &--\\
  3 & $x^2$ & $2/3$ &-- \\
  4 & $x^3$ & $1/2$ &--\\
  5 & $x^4$ & $2/5$ &--\\
\midrule
  6 & $1-x$& 1 &--\\
  7 & $1-x^2$& $4/3$ &\cite{M7C1, M7C2}\\
  8 & $1-x^3$&  $3/2$&\cite{M7C1, M7C2}\\
  9 & $1 -\frac{3 x}{2}+\frac{x^3}{2}  $ & $3/4$ & \cite{MY08}\\
 10 & $(1-x)^2$&  $2/3$  & \cite{MY08} \\
 11 & $(1-x)^3$& 1/2  &--\\
 12 & $(1-x)^2 (1+2 x)$ & 1 & $w_\tau$  \\
 13 & $(1-x)^3 (1+2 x)$ & 7/10&\cite{Davier} \\
\midrule
 14 & $(1-x)^2 x $ & 1/6 & \cite{MY08}\\
 15 & $(1-x)^3 x (1+2 x)$ &1/6 & \cite{Davier} \\
 16 & $(1-x)^3 x^2 (1+2 x)$& 13/210 &\cite{Davier}  \\
 17 & $(1-x)^3 x^3 (1+2 x)$ & 1/35 &\cite{Davier} \\
  \bottomrule
\end{tabular}
\caption{Weight functions investigated in this analysis, together with the
corresponding $\delta^{\rm tree}_{w_i}$. In the last column, we give the
reference to recent works that employed the given weight function in analyses
of $\tau$ decay data. The kinematical weight function $w_\tau$ was used many
times throughout the literature and we refrain from quoting all the works that
employed it.\label{tab:ws}}
\end{center}
\end{table}

In the remainder of this section we study the behaviour of the term
$\delta^{(0)}_{w_i}(m_\tau^ 2)$, for the moments of table~\ref{tab:ws} in
FOPT and CIPT, given by eqs.~(\ref{del0FO}) and~(\ref{del0CI}) respectively.
In doing so, we employ for the coefficient $c_{5,1}$ the estimate of
eq.~(\ref{c51}). Regarding higher orders, two scenarios are considered: the
reference model of ref.~\cite{BJ2008}, given in eq.~(\ref{CMBJ}), and the
alternative model of eq.~(\ref{ArtifModel}), which provides an example case
where CIPT is better than FOPT for the kinematical weight.  In plots for the
perturbative series we display the results for both models side by side to
facilitate the comparison. The respective Borel resummed values are also shown,
together with the Borel ambiguity.  Since both models are matched to the
first five coefficients, their results for CIPT and FOPT are identical by
construction up to the fifth order.
 
In our analysis, it becomes clear that one can group the 17 different
$\delta^{(0)}_{w_i}$ into four classes: the monomial terms, the pinched-weights
with a ``1'' in the weight function, pinched weights without a ``1'', and all
moments that contain the term $x$, which form a separate category. We analyse
these classes in the remainder of this section.

\subsection{Building blocks: the monomial terms}

Since the weight functions employed are all polynomial, it is instrumental to
start the analysis with the monomial terms given in the first 5 entries of
table~\ref{tab:ws}. 

\begin{figure}[!ht]
\begin{center}
\subfigure[$w_1=1$, reference model]{\includegraphics[width=.48\columnwidth,angle=0]{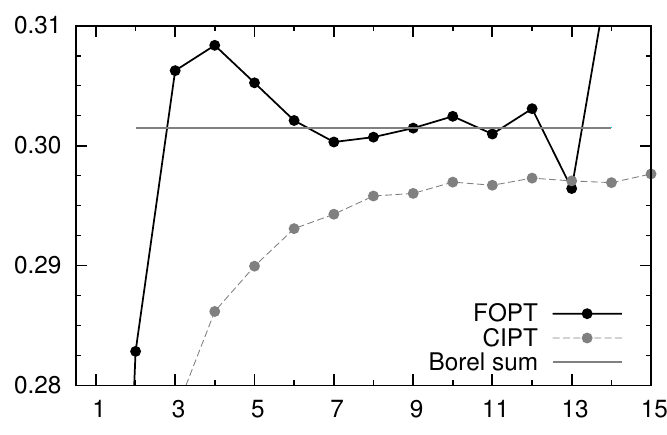}\label{w1PM}}
\subfigure[$w_1=1$, alternative model]{\includegraphics[width=.48\columnwidth,angle=0]{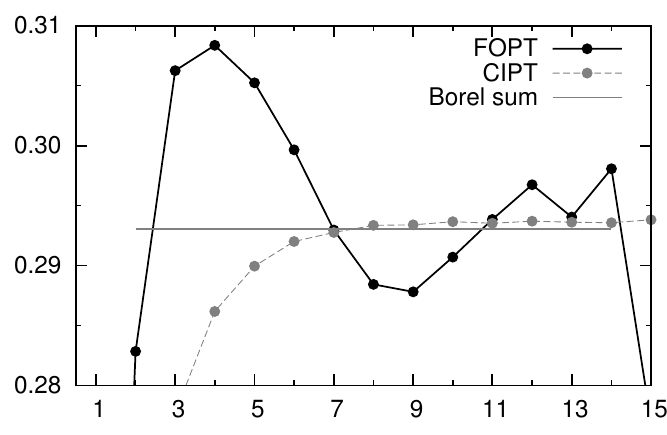}\label{w1AM}}
\subfigure[$w_2=x$, reference model]{\includegraphics[width=.48\columnwidth,angle=0]{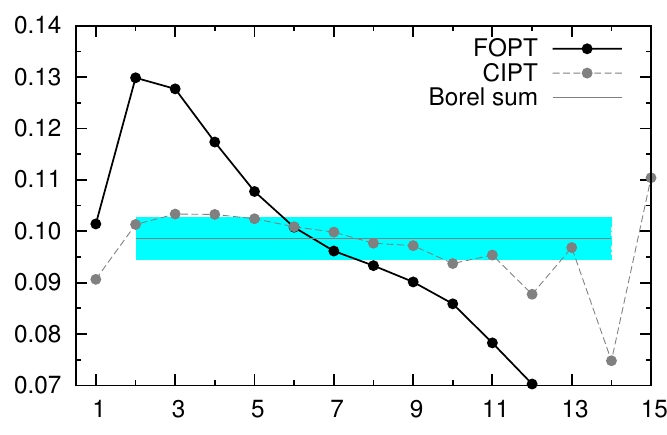}\label{wxPM}}
\subfigure[$w_2=x$, alternative model]{\includegraphics[width=.48\columnwidth,angle=0]{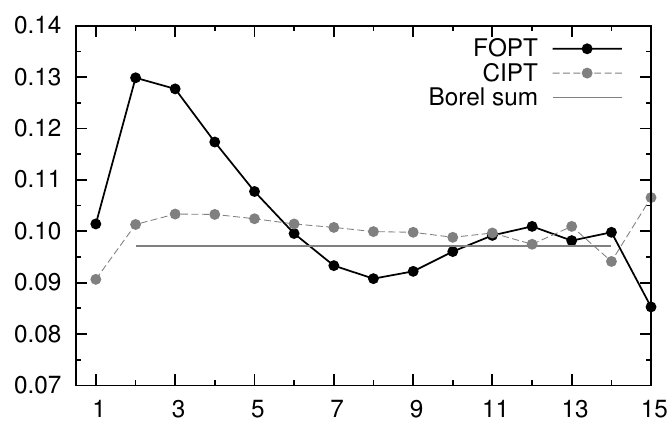}\label{wxAM}}
\subfigure[$w_3=x^2$, reference model]{\includegraphics[width=.48\columnwidth,angle=0]{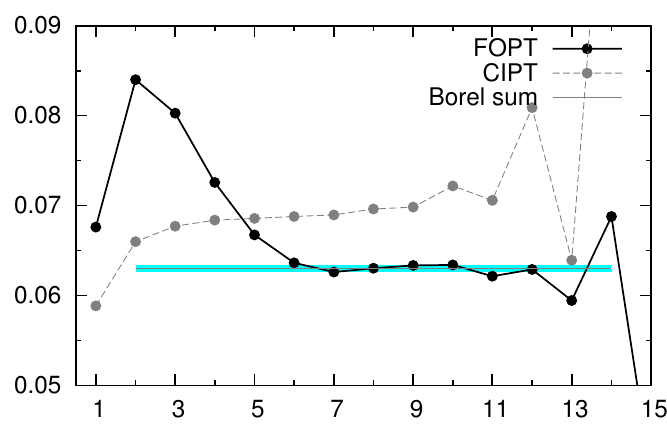}\label{wxsqPM}}
\subfigure[$w_3=x^2$, alternative model]{\includegraphics[width=.48\columnwidth,angle=0]{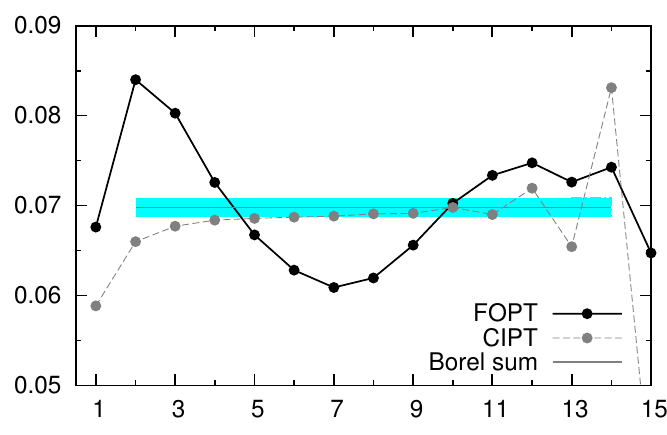}\label{wxsqAM}}
\caption{$\delta^{(0)}_{w_i}$ for $w_1=1$, $w_2=x$, and $w_3=x^2$, as a
function of the order up to which the perturbative series are summed for FOPT
(black) and CIPT (gray). The horizontal bands give the Borel resummed result.
The left-hand figures are for the RM of ref.~\cite{BJ2008}, the right-hand
ones are for the alternative model of eq.~(\ref{ArtifModel}). We use
$\alpha_s(m_\tau)=0.3186$.\label{BuildingBlocks}}
\end{center}     
\end{figure}

The first of them, the constant $w_1=1$, is quite particular. On the
experimental side this moment gives the same weight to the whole spectrum. On
the theory side, it obtains very little contribution from condensates, since
no powers of $x$ appear, picking up only the logarithmic contributions in the
Wilson coefficients of the OPE.\footnote{This is precisely the reason why this
moment is central to the analysis of refs.~\cite{M7C1,M7C2}, where one wants to
extract the DVs from data.}  The behaviour of $\delta^{(0)}_{w_1}$ for FOPT
and CIPT is shown in figure~\ref{w1PM} for the reference model, and in
figure~\ref{w1AM} for the alternative model. A feature that can be observed in
FO is a rapid growth in the first few terms, followed by a decrease in the
value of $\delta^{(0)}_{w_1}$. In lower orders, the series overshoots the Borel
summed values for both models. Later, FOPT oscillates around the Borel sum. The
amplitude of the oscillations are much smaller in the RM, and from order $n=5$ 
the series can be considered a good approximation to the true result in the
sense of an asymptotic series. The smaller oscillations in the RM model are due
to the previously mentioned cancellations among the $c_n$ and the $g_n^{[w_1]}$
series. CIPT is more stable in both cases, but it fails to give a good
approximation to the Borel resummed value of the RM. In the case of the
alternative model, CIPT is able to approach the true value, as expected. The
smallness of the ambiguity of the Borel integral due to the poles on the
integration contour (indicated by the horizontal shaded band in the figures) in
these cases can be understood since the moment receives only small logarithmic
contributions from the condensates in the OPE (for lack of powers of $x$).
Accordingly, the ambiguities of the IR poles are also small.\footnote{They
would be zero if the poles were simple. Since we include the four-loop
structure for the renormalon singularities, they are small, but non-zero. See
Appendix~A of ref.~\cite{BJ2008} for the explicit formulae.}

The behaviour of $\delta^{(0)}_{w_i}$ for the monomials $w_3=x^2$, $w_4=x^3$,
$w_5=x^4$ is qualitatively very similar. Therefore, we display only the
representative case of $x^2$ in figures~\ref{wxsqPM} and~\ref{wxsqAM} for the
RM and AM, respectively, which highlights these similarities. We note that the
values of $\delta^{(0)}_{w_i}$ are $4$ to $6$ times smaller than the ones for
$w_1$. This plays an important role in the case of moments with pinching.
Finally, $w_3$ is maximally sensitive to the Borel ambiguity of the IR
singularity at $u=3$, present in both models. However, the residue in the case
of the RM is about 5 times smaller than in the AM, which explains the different
magnitudes of the shaded bands in the two plots. 

The behaviour of the monomial $w_2=x$ is exceptional, as shown in
figure~\ref{wxPM} for the reference model, and in figure~\ref{wxAM} for the
alternative model. One can separate the behaviour in two parts. In the first
terms, the FOPT series again grows rapidly and then decreases. This is a common
feature in the other monomials as well. For higher orders, in the RM, FOPT
never reaches a plateau: $\delta^{(0)}_{w_2}$ exhibits ``run-away'' behaviour
and decreases monotonically from the 3rd order. The FOPT series shows no sign
of stabilisation around the true value, though it develops an inflection point
close to the Borel sum. The sign of the run-away behaviour (negative) is
correlated with the sign of the $x$ monomial (positive). For the alternative
model, FOPT still oscillates around the true value. CIPT, on the other hand,
is rather stable until the onset of asymptoticity, and provides a good
approximation in both models. The results in this prescription are within the
Borel resummed values for the RM, given the larger ambiguity in this case. The
larger ambiguity stems from the fact that the moment is maximally sensitive to
the gluon condensate contribution. This gives a larger contribution from the
ambiguity of the IR singularity at $u=2$, which is (artificially) absent in
the AM. 

From this discussion of monomial moments we can already extract a few important
observations:
\begin{itemize}
\item In the AM CIPT always provides the better approximation. This seems to
be a generic feature of the AM, not restricted to the kinematical weight.
\item In the reference model, the monomial $w_2=x$ is problematic for FOPT due
to run-away behaviour, which is correlated with the large $d=4$ condensate
contribution to this moment. 
\item For the other monomials low-order approximations in both FOPT and CIPT
are problematic in the RM case. However, while FOPT converges to the Borel sum
for $n \gsim 5$, CIPT never reaches it. This is similar to the behaviour found
in ref.~\cite{BJ2008} for the kinematical weight.
\end{itemize}
In the following, we discuss the remaining moments $w_6$ to $w_{17}$ which are
composed of the monomial terms. Their behaviour can essentially be understood
as a linear combination of what has been discussed in this section.

\subsection{Pinched weights with a ``1''}

We now turn our attention to moments with pinching and start with moments that
contain a term ``1'' in the weight function and that do not have a linear term
$x$. In table~\ref{tab:ws}, these moments are $w_7$, $w_8$, $w_\tau$.  For all
these moments, the term $1$ sets the scale and the higher powers only introduce
corrections to this leading result. Since they are pinched moments, they always
have at least one negative term. This leads, in general, to a stabilisation of
FOPT in both models.

\begin{figure}[!ht]
\begin{center}
\subfigure[$w_\tau=(1-x)^2 (1+2 x)$, reference model]{\includegraphics[width=.48\columnwidth,angle=0]{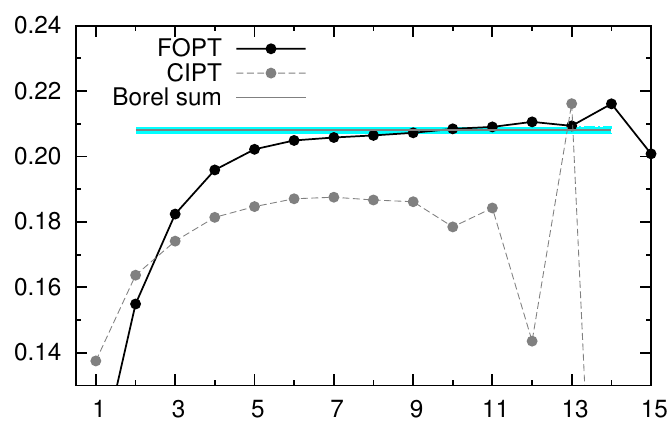}\label{wtauCM}}
\subfigure[$w_\tau=(1-x)^2 (1+2 x)$, alternative model]{\includegraphics[width=.48\columnwidth,angle=0]{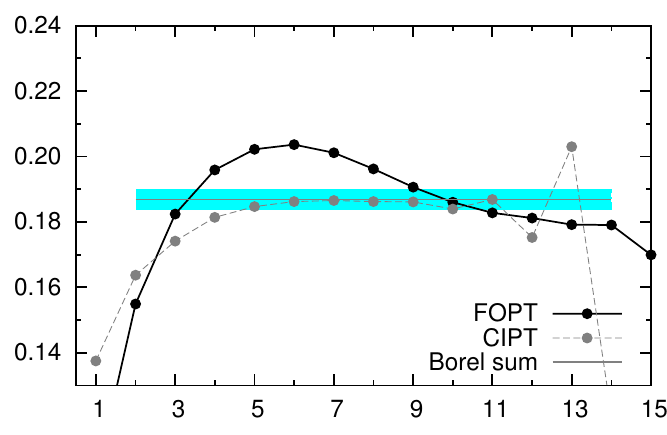}\label{wtauAM}}
\subfigure[$w_7=1-x^2$, reference model]{\includegraphics[width=.48\columnwidth,angle=0]{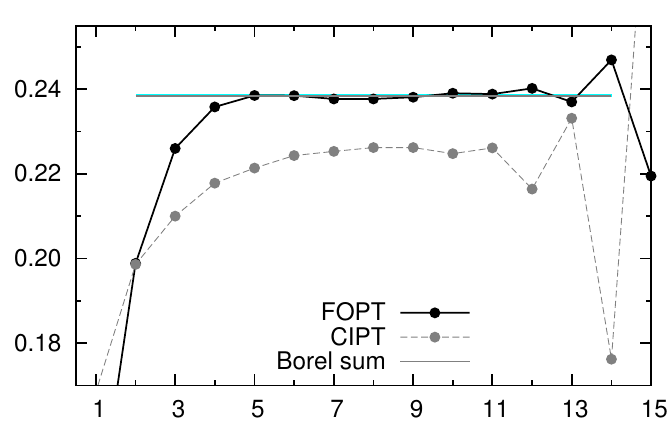}\label{w7CM}}
\subfigure[$w_7=1-x^2$, alternative model]{\includegraphics[width=.48\columnwidth,angle=0]{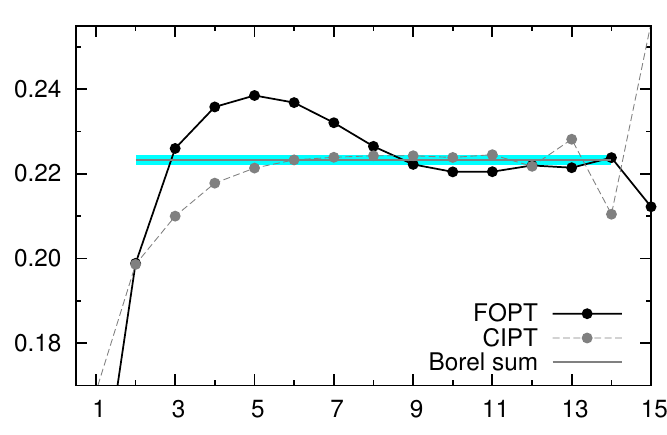}\label{w7AM}}
\subfigure[$w_8=1-x^3$, reference model]{\includegraphics[width=.48\columnwidth,angle=0]{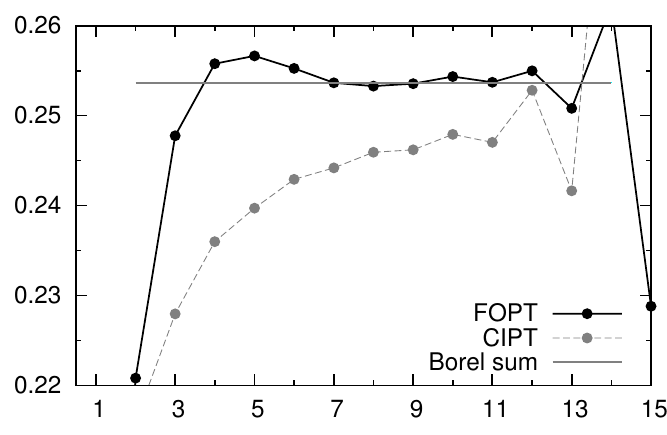}\label{w8CM}}
\subfigure[$w_8=1-x^3$, alternative model]{\includegraphics[width=.48\columnwidth,angle=0]{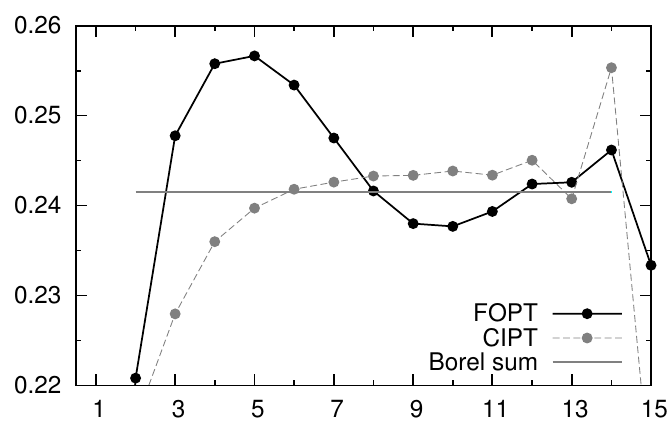}\label{w8AM}}
\caption{$\delta^{(0)}_{w_i}$ for $w_\tau$, $w_7$, and $w_8$, as a function of
the order up to which the perturbative series are summed for FOPT (black) and
CIPT (gray). The horizontal bands give the Borel resummed result. The left-hand
figures are for the RM of ref.~\cite{BJ2008}, the right-hand ones are for the
alternative model of eq.~(\ref{ArtifModel}). We use $\alpha_s(m_\tau)=0.3186$.
\label{Pinched1}}
\end{center}
\end{figure}

In the case of FOPT for the RM, the monomials in $w_\tau$ and $w_7$ conspire
to give an excellent cancellation of the initial overshooting, leading to a
series that approaches the Borel sum very fast. The cancellations between
$c_n\delta_{w_i}^{\rm tree}$ and $g_n^{[w_i]}$ then make FOPT rather stable
after approaching the true value. This can be observed in the results for the
RM given in figures~\ref{wtauCM}, \ref{w7CM}, and \ref{w8CM}. The results for
$w_8=1-x^3$ resemble more the ones for $w_1=1$ since the corrections due to
$x^3$ are quite small. CIPT, on the other hand, misses the cancellations and
never approaches the Borel resummed values. We also observe that the
CIPT series enters the sign alternating regime earlier than FOPT. The
corresponding results for the AM are shown in figures~\ref{wtauAM}, \ref{w7AM},
and \ref{w8AM}. As foreseen, here CIPT tends to give a better approximation to
the resummed series. Albeit more stable than in the case of the monomials,
FOPT still displays oscillations around the Borel sum.

There are many other possible moments, not shown in table~\ref{tab:ws}, that
display a very similar behaviour. We have investigated the family
$w(x;a)=1+ax^2-(a+1)x^{3}$ for several different values of $a$ with results
rather similar to the ones discussed above. Other moments that do not start
with the unity, but with another constant of the same order such as
$w(x)=\frac{2}{3}(1-x)^2(1+x)(1+x+4x^2)$ also give qualitatively similar
results. Thus, the main observation for this class is that
\begin{itemize}
\item pinched-moment weights with a ``1'' but no linear term all behave similar
to the kinematic weight $w_\tau$, favouring FOPT over CIPT for the reference
model and vice versa for the alternative model. In each case the perturbative
expansion in low orders approaches the Borel sum rather quickly.
\end{itemize}

\subsection{Pinched weights without a ``1''}

The next class of moments that we analyse here are moments with pinching but
that do neither contain a constant term nor one linear in $x$. As examples for
this group we employ $w_{16}$ and $w_{17}$ of table~\ref{Moments}; two of the
triply-pinched moments used by ALEPH and OPAL~\cite{ALEPH,ALEPH2, Davier, OPAL}. In ALEPH's
notation, these moments are part of a family denoted by $w^{(1,k)}$, and read
\begin{equation}
\label{ALEPHMom}
w^{(1,k)} \,=\, (1-x)^3 x^k (1+2x) \,=\,
x^k - x^{k+1} - 3\,x^{k+2} + 5\,x^{k+3} - 2\,x^{k+4} \,.
\end{equation}
In table~\ref{Moments} we have $w_{13}=w^{(1,0)}$, $w_{15}=w^{(1,1)}$,
$w_{16}=w^{(1,2)}$, and $w_{17}=w^{(1,3)}$. Besides the ``1'' in $w_{13}$,
both the first two also contain a linear term in $x$ and will be discussed
in the next section. From table~\ref{tab:ws} we see that the pinched weights
without a ``1''  have very small $\delta_{w_i}^{\rm tree}$ but ${\cal O}(1)$
coefficients of the monomials. This enhances the relative importance of power
corrections.

\begin{figure}[!ht]
\begin{center}
\subfigure[$w_{16}=(1-x)^3 x^2 (1+2 x)$, reference model]{\includegraphics[width=.45\columnwidth,angle=0]{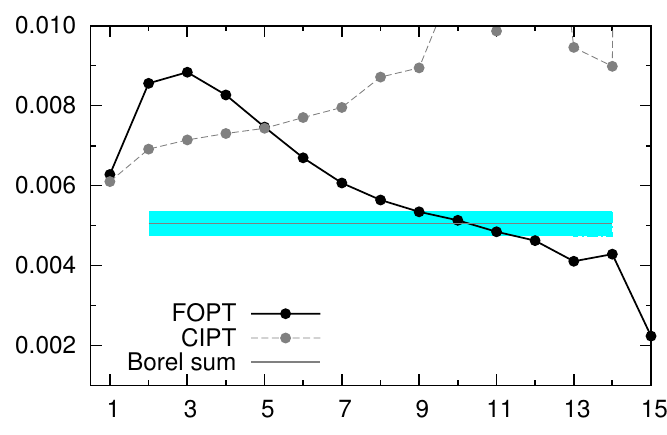}\label{w16PM}}
\subfigure[$w_{16}=(1-x)^3 x^2 (1+2 x)$, alternative model]{\includegraphics[width=.45\columnwidth,angle=0]{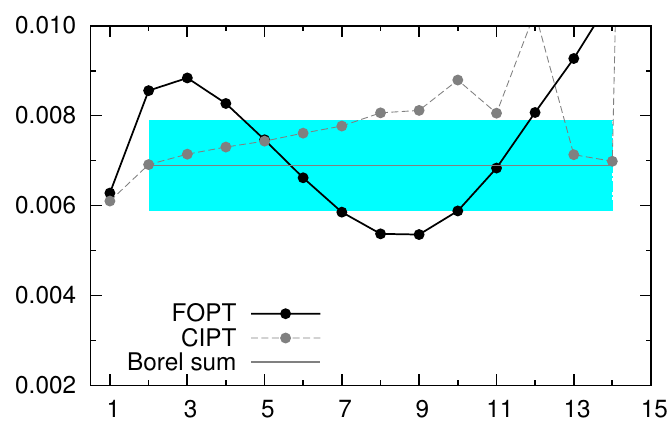}\label{w16AM}}
\subfigure[$w_{17}=(1-x)^3 x^3 (1+2 x)$, reference model]{\includegraphics[width=.45\columnwidth,angle=0]{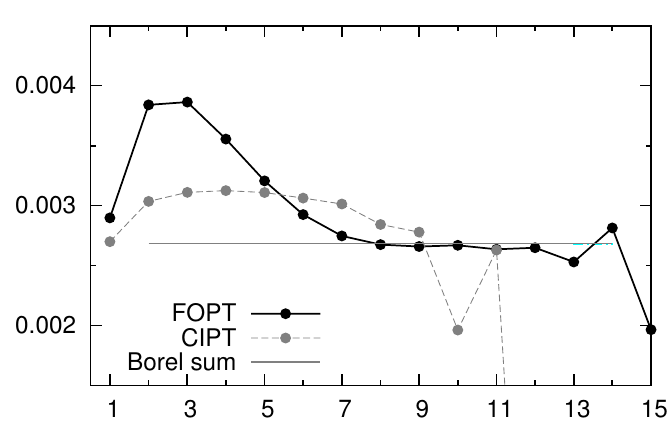}\label{w17PM}}
\subfigure[$w_{17}=(1-x)^3 x^3 (1+2 x)$, alternative model]{\includegraphics[width=.45\columnwidth,angle=0]{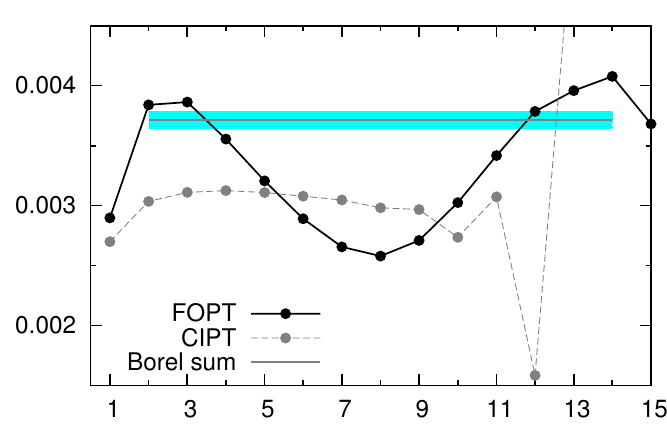}\label{w17AM}}
\caption{$\delta^{(0)}_{w_i}$ for $w_{16}$ and $w_{17}$, as a function of the
order up to which the perturbative series are summed for FOPT (black) and CIPT
(gray) are summed. The horizontal bands give the Borel resummed result. The
left-hand figures are for the RM of ref.~\cite{BJ2008}, the right-hand ones are
for the alternative model of eq.~(\ref{ArtifModel}). We use
$\alpha_s(m_\tau)=0.3186$.
\label{PinchedX}}
\end{center}     
\end{figure}

We consider first the case of $w_{16}$ in the RM, shown in figure~\ref{w16PM}.
As expected, $\delta^{(0)}_{w_{16}}$ is tiny, almost 50 times smaller than the
corresponding correction for $w_1=1$. The combination of powers of $x$ does
not improve the behaviour of FOPT that overshoots largely the Borel sum in the
first few orders. It eventually approaches the Borel result for higher orders,
just before the onset of asymptoticity. The bad behaviour of CIPT already
observed for the monomials is amplified and the CIPT series goes astray. The
situation in the AM is somewhat improved, but mainly due to the large Borel
ambiguity associated with the IR pole at $u=2$, see figure~\ref{w16AM}. Again,
FOPT displays large oscillations around the Borel result, whereas CIPT grows
monotonically away from the resummed result before the sign-alternating
asymptotic behaviour sets in. As seen in figure~\ref{w17PM}, for FOPT in the
RM, the perturbative contribution to the moment of $w_{17}$ has a behaviour
qualitatively similar to $w_{16}$, though for higher orders it is slightly more
stable. Also CIPT approaches the Borel sum before the series becomes asymptotic
after the 9th order. In the AM, figure~\ref{w17AM}, both, FOPT and CIPT fail to
approach the Borel sum, FOPT once more displaying large oscillations.\footnote{
A possible criticism against our analysis of $w_{17}$ within the reference
model could regard the lack of an IR singularity at $u=4$. Since the moment
starts with $x^3$ it is maximally sensitive to $D=8$ contributions in the
OPE, which corresponds to the ambiguity of the IR renormalon at $u=4$. We
investigated this issue by considering a model where one adds an IR renormalon
at $u=4$ and leaves only a constant $d_{0}^{\rm PO}$ in the model. (This
model is briefly discussed on page~24 of ref.~\cite{BJ2008}.) After performing
the matching, the residue of the renormalon at $u=4$ turns out to be small
($d_4^{\rm IR}=5.64$) and the changes in the other parameters negligible. The
additional Borel ambiguity arising from $u=4$ is also small. Therefore, the
result shown in figure~\ref{w17PM} is not altered in any significant way,
which corroborates the assumption that the singularities at $u=2$ and $u=3$
are the dominant ones.} We therefore conclude:
\begin{itemize}
\item The perturbative expansions for this class of moments tends to be 
unreliable in both FOPT and CIPT, and independent of the model for the unknown
higher-order coefficients.
\item Pinched moments without a ``1'' are sensitive to condensates, but the
poor perturbative approximations render condensate determinations from these
moments unreliable. This conclusion appears to be largely model-independent.
\end{itemize}

\begin{boldmath}
\subsection{Moments containing a term $x$}
\end{boldmath}

We have relegated to this section the analysis of weight functions containing
the monomial $x$. This choice is based on the observation that the behaviour of
these moments, which are maximally sensitive to the $D=4$ correction in the OPE,
is qualitatively different in the RM of ref.~\cite{BJ2008}. This was shown for
the monomial above and here we discuss pinched weights containing this term.
There are several of them in table~\ref{Moments}: $w_6$, $w_9$, $w_{10}$,
$w_{11}$, $w_{13}$, $w_{14}$, and $w_{15}$. Again, they display very similar
qualitative behaviours and it suffices to expose in detail only three
representative examples.
 
\begin{figure}[!ht]
\begin{center}
\subfigure[$w_6=1-x$, reference model]{\includegraphics[width=.48\columnwidth,angle=0]{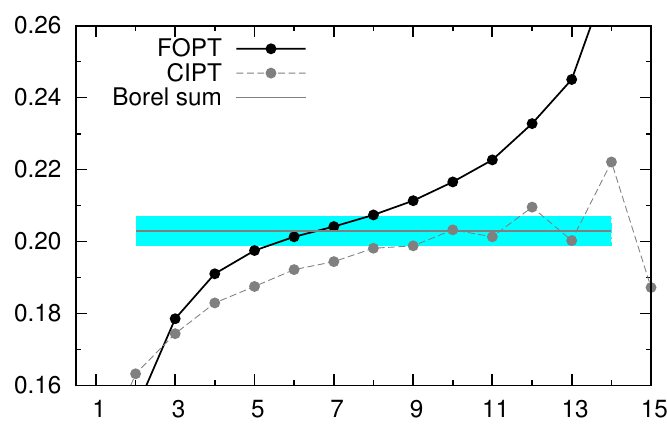}\label{w6PM}}
\subfigure[$w_6=1-x$, alternative model]{\includegraphics[width=.48\columnwidth,angle=0]{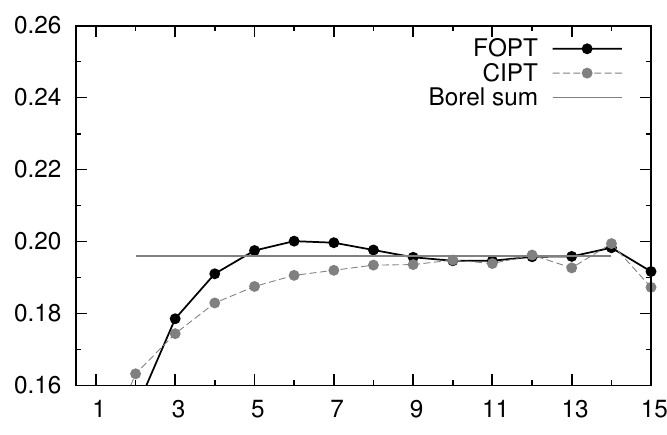}\label{w6AM}}
\subfigure[$w_{13}=(1-x)^3 (1+2 x)$, reference model]{\includegraphics[width=.48\columnwidth,angle=0]{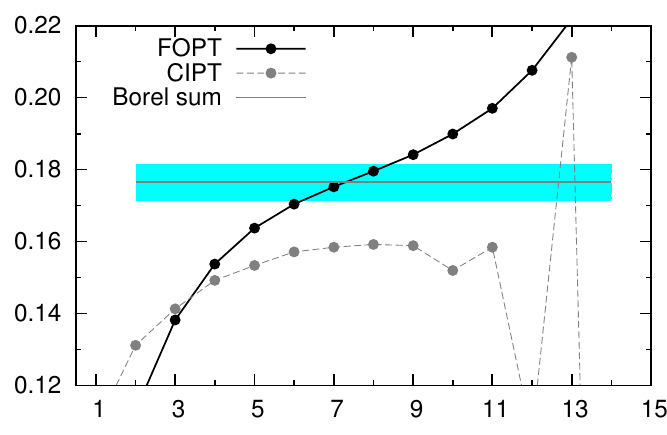}\label{w13PM}}
\subfigure[$w_{13}=(1-x)^3 (1+2 x)$, alternative model]{\includegraphics[width=.48\columnwidth,angle=0]{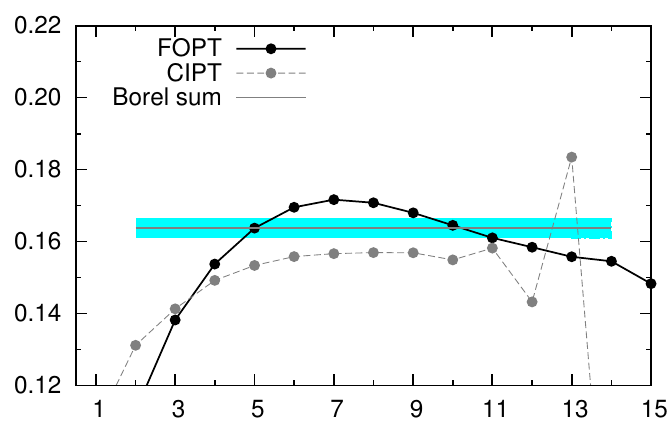}\label{w13AM}}
\subfigure[$w_{15}=(1-x)^3 x (1+2 x)$, reference model]{\includegraphics[width=.48\columnwidth,angle=0]{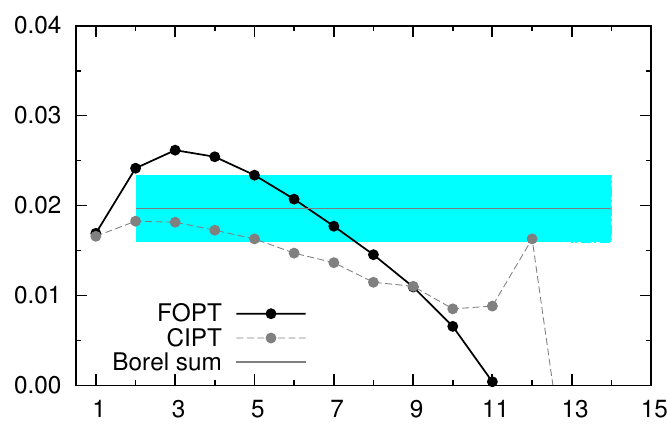}\label{w15PM}}
\subfigure[$w_{15}=(1-x)^3 x (1+2 x)$, alternative model]{\includegraphics[width=.48\columnwidth,angle=0]{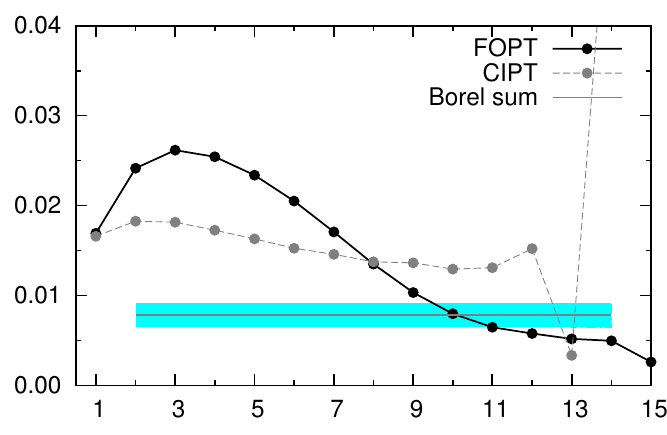}\label{w15AM}}
\caption{$\delta^{(0)}_{w_i}$ for $w_6$, $w_{13}$, and $w_{15}$, as a function
of the order up to which the perturbative series are summed for FOPT (black)
and CIPT (gray). The horizontal bands give the Borel resummed result. The
left-hand figures are for the RM of ref.~\cite{BJ2008}, the right-hand ones
are for the alternative model of eq.~(\ref{ArtifModel}). We use
$\alpha_s(m_\tau)=0.3186$.
\label{XMoments}}
\end{center}     
\end{figure}

We start with the simple case of $w_6=1-x$, figure~\ref{w6PM}. The term $1$
sets the scale, but now, since the perturbative series for the monomial $w_2=x$
decreases monotonically, the perturbative series for $w_6$, whose linear
coefficient has negative sign,  grows monotonically. The result for FOPT
crosses the Borel resummed value around the 7th order where it also has an
inflection point. CIPT can only approach the Borel sum shortly before it
becomes asymptotic around the 9th order. The situation is very similar when
higher orders of $x$ are added to the weight function. In figure~\ref{w13PM},
the moment $w_{13}$ is displayed as a representative example. (It corresponds
to $w^{(1,0)}$ in ALEPH's notation~\cite{ALEPH,ALEPH2,Davier} of eq.~(\ref{ALEPHMom})).
This weight starts with $1-x$ followed by higher-order terms in $x$.
Qualitatively, the only difference now is that CIPT never comes close to the
Borel sum. The picture changes in the case of $w_{15}$ (or $w^{(1,1)}$) since
the term $1$ is missing. As figure~\ref{w15PM} shows, this case is more
similar to the monomial $x$ itself, though the higher powers in $x$ soften
the behaviour. For CIPT, the situation is not much different either, but it
only approximately represents the Borel sum  at low orders.

For the alternative model which does not contain the renormalon singularity at
$u=2$, the first two moments $w_6$ and $w_{13}$, figures~\ref{w6AM} and
\ref{w13AM}, are similar to what was observed in the case of pinched moments
with the term 1. Now, however, CIPT approaches the Borel results less fast in
the case of $w_6$, and not at all in the case of $w_{13}$. FOPT in both cases
displays oscillations around the true value. The last case, that of $w_{15}$
shown in figure~\ref{w15AM}, is less satisfactory. As was the case for other
moments from weight functions starting with a power of $x$, the values of
$\delta^{(0)}_{w_{15}}$ are small. While CIPT misses the Borel sum completely,
FOPT only approaches it around the 10th order. We summarise our observations
as:\\[-0.2cm]

\begin{itemize}
\item Weights without a ``1'' are again unreliable in FOPT and CIPT, and in
both models, especially at intermediate orders.
\item In the RM, FOPT exhibits run-away behaviour and CIPT may not approach
the resummed result. Overall, the perturbative expansion does not behave as
well as for moments without a linear term, which is related to the sizeable
$D=4$ power correction.
\item In the AM there is no clear preference for one of the two methods. 
\end{itemize}

\subsection{Main lessons from the moment analysis}

While we already summarised our main observations for each class 
of weight functions, we collect again here the most important points.

Some of the pinched-moments (with the ``1'', without the ``$x$'') display a
particularly fast convergence of FOPT towards the Borel resummed values,
especially the moments $w_\tau$ and $w_7$, as shown in figures~\ref{wtauCM}
and~\ref{w7CM}. On the contrary, in Borel models that contain a $u=2$ pole
residue of natural size, CIPT generally does not approach the Borel sum before
the divergence of the series  sets in. This is different in the alternative
model, where the $u=2$ pole is artificially suppressed, and it coincides with
the main findings of  ref.~\cite{BJ2008}. Thus, if the reference model is
adopted as the most plausible one (as we would do), one again arrives at the
conclusion that FOPT provides a better approximation than CIPT, also at order
$n=4,5$.

The investigation of moments also reveals that the qualitative behaviour of
their perturbative expansion depends only on a few features of the moment
and the model of the Adler function. As concerns the model, we have already
emphasised the crucial question of the size of the residue of the $u=2$
singularity that corresponds to the $D=4$ power correction, which motivated
the choice of the AM.

As concerns the moment function itself,  an important observation is that
moments that start with high powers of $x$, such as $w_{15}$ and $w_{16}$,
employed by the ALEPH and OPAL collaborations, have a bad behaviour of the
perturbative series. For these moments, neither FOPT nor CIPT are able to
provide a decent approximation to the Borel resummed values in the first few
orders. Also, these moments have a very small value of $\delta^{(0)}_{w_i}$,
with large relative Borel ambiguities, which makes the reliable separation of
power corrections from the uncertain perturbative approximation problematic.
Therefore, these moments are not an optimal choice for an $\alpha_s$ analysis.
(This has already been pointed out in ref.~\cite{MY08}.)

Finally, for moments that contain a linear term $x$, the reference model, or
others that include an IR pole at $u=2$, both FOPT and CIPT behave badly. In
the case of FOPT, the series is quite unstable, which results in large errors
due to the truncation of the series, producing unstable results for $\alpha_s$.
(This was noticed --- in practice --- in the exploratory fits of
ref.~\cite{Manchester}, and it is the reason why, in refs.~\cite{M7C1,M7C2},
moments with the term $x$ were not considered.) The situation of CIPT for
these moments is also unsatisfactory because the series are unstable and/or
do not approach the Borel resummed result. This suggests that moments with a
linear term $x$ should also be avoided in $\alpha_s$ determinations.

\section{Validation of the reference  model }
\label{Plausibility}

In the previous section we learnt that for moments with good perturbative 
convergence the comparison of FOPT and CIPT leads to the same conclusion as
for the inclusive hadronic tau width studied in ref.~\cite{BJ2008}. Hence,
the crucial factor in deciding whether FOPT or CIPT should be the method of
choice remains the plausibility of the reference ansatz for the Adler function
(favouring FOPT) as compared to, e.g., the alternative model (favouring CIPT).
 In addition to the general arguments for the reference model reviewed in 
section~\ref{RenormalonModels}, we discuss in this section two further checks,
one inspired by ref.~\cite{DGM}, which support the plausibility of the ansatz
and results of  ref.~\cite{BJ2008}.

\begin{boldmath}
\subsection{Adding a $u^2$ polynomial term}
\end{boldmath}

In ref.~\cite{BJ2008}, the known higher-order coefficients, plus an estimate
for $c_{5,1}$, are used to fix the residua of the renormalon poles, while the
first two polynomial terms are obtained by also fitting the coefficients
$c_{1,1}$ and $c_{2,1}$. It is assumed, therefore, that the renormalons
dominate at intermediate (and higher) perturbative orders. However, the fact
that $d_1^{\rm PO}$ is small in the reference model of~\cite{BJ2008},
indicates that $c_{2,1}$ is already well saturated by the renormalon poles.
The procedure has been criticised in ref.~\cite{DGM}, where the authors argue
that the truncation of the polynomial terms at linear order is arbitrary. They
propose to add a $u^2$ term to the polynomial and study the behaviour of those
models when the coefficient $d_2^{\rm PO}$ is fixed to six different values:
$d_{2}^{\rm PO} = -1,\,-0.5,\,0,\,0.25,\,0.5,\,1$. For the value
$d_{2}^{\rm PO} = 0$, the RM is recovered.

The inclusion of a {\it fixed} $u^{2}$ term in the modelling of the Borel
transform of the Adler function has consequences for the residua of the
renormalon poles. They have to adjust to the existence of this term which
contributes to $c_{3,1}$, which leads to abnormally high values for the residue
of the IR pole at $u=3$, see table~\ref{CMvsDGM}. Consequently, large
cancellations among the contributions of the IR poles at $u=2$ and $u=3$ arise.
In the two extreme cases studied in~\cite{DGM} , $d_2^{\rm PO}=\pm 1$, the
residue of the pole at $u=3$ changes by factors of $-11$ and $13$ with respect
to the RM of~\cite{BJ2008}, for $d_2^{\rm PO}=+1$ and $d_2^{\rm PO}=-1$
respectively. Furthermore, fixing the $u^2$ term also forces a break-down of
the renormalon dominance of the coefficients $c_{2,1}$ and $c_{3,1}$. This is
apparent from the values of the other polynomial terms. In the extreme cases
$d_2^{\rm PO}=\pm 1$, the module of the coefficient $d_0^{\rm PO}$ is more than
10 times larger than in the RM. The next coefficient, $d_1^{\rm PO}$, is also
much larger, more than 100 times the one found in~\cite{BJ2008}. The values for
the residua and the polynomial terms for the RM and for the extreme cases
$d_2^{\rm PO}=\pm 1$ are given in table~\ref{CMvsDGM}. Finally, the coefficient
$d_2^{\rm PO}$ and the $u=2$ residue $d_2^{\rm IR}$ share an almost linear
relation, such that $d_2^{\rm IR}$ vanishes for $d_2^{\rm PO}=0.678$. This
particular case constitutes another model for which CIPT generally better
approximates the Borel sum for $\delta^{(0)}_{w_i}$.

\begin{table}[t]
\begin{center}
\begin{tabular}{r r r r r c r}
\toprule
& $d_2^{\rm IR}$ & $d_3^{\rm IR}$ & $d_{1}^{\rm UV}\qquad$ & $d_0^{\rm PO}$& $d_1^{\rm PO}$ & $d_2^{\rm PO}$ (fixed) \\
 \midrule
RM~\cite{BJ2008}   &  $3.16$ & $-13.5$ &  $-1.56\times 10^{-2}$ & $0.78$  & $7.66\times 10^{-3}$ & $0\qquad$ \\
$+u^ 2$~\cite{DGM} & $-1.50$ & $149.7$ &  $-5.90\times 10^{-2}$ & $10.68$ & $\phantom{-}3.85$ & $1\qquad$ \\
$-u^ 2$~\cite{DGM} &  $7.82$ & $-176.8$ & $2.79\times 10^{-2}$  & $-9.12$ & $-3.83$ &  $-1\qquad$ \\
 \bottomrule
\end{tabular}
\caption{Parameter values of three models for the physical Adler function.
``RM'' represents the central reference model of ref.~\cite{BJ2008}. The models
denoted by ``$\pm u^2$'' are those discussed in ref.~\cite{DGM} where the
polynomial coefficient $d_2^{\rm PO}$ is taken to be $\pm 1$.
\label{CMvsDGM}}
\end{center}
\end{table}

\begin{figure}[!ht]
\begin{center}
\includegraphics[width=12cm,angle=0]{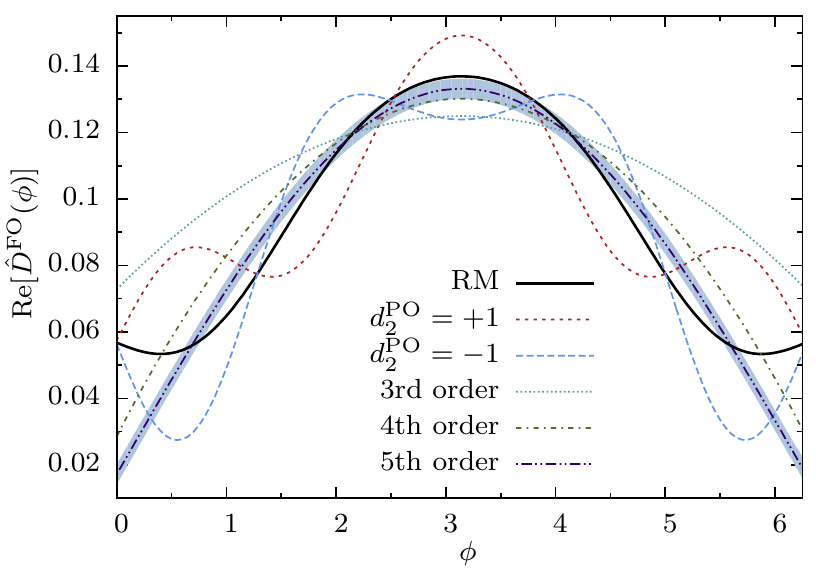}
\includegraphics[width=12cm,angle=0]{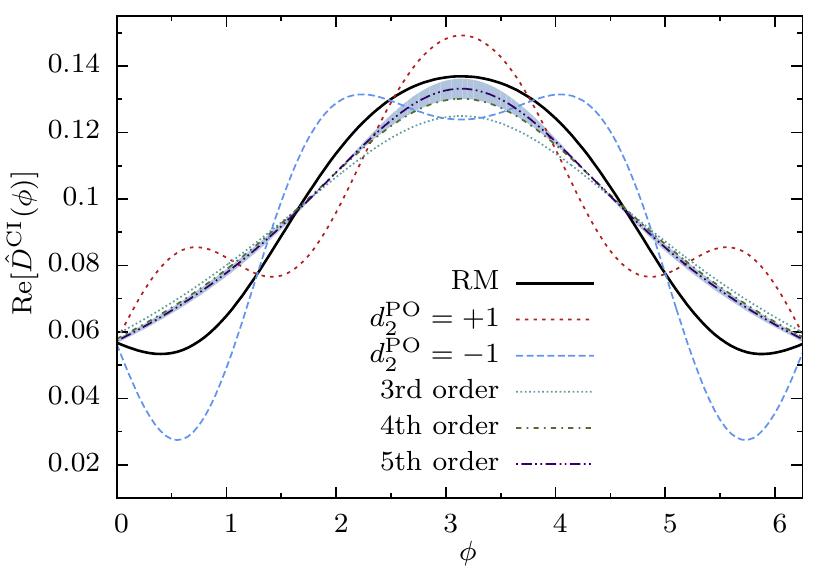}
\caption{ Real part of the Adler function $\widehat D(s)$ on the complex
circle $s=m_\tau^ 2 e^{i\phi}$. Borel sums are displayed for the RM (solid
line), $d_2^{\rm PO}=1$ (short-dashed line) and $d_2^{\rm PO}=-1$ (long-dashed
line).  The dotted, dot-dashed and double-dot-dashed lines correspond to 3rd,
4th and 5th order of perturbation theory. Finally, the shaded area indicates
the 5th order PT result while varying $c_{5,1}=283\pm283$. The upper plot
shows FOPT and the lower CIPT. We employ $\alpha_s(m_\tau)=0.3186$.}
\label{FOAndCIAdlerFunctiond2models}
\end{center}
\end{figure}

The inspection of the Adler function $\widehat D(s)$ on the complex circle
$s=m_\tau^2 {\rm e}^{i\phi}$ sheds further light on the plausibility of the
models considered here. It is expected that the perturbative expansion breaks
down in the vicinity of the physical, Minkowskian axis ($\phi\sim 0$ or
$\phi\sim 2\pi$), but that it should work well in the Euclidean region
$\phi\sim \pi$. The behaviour of $\Re[\widehat D(\phi)]$ along the complex
contour is displayed in figure~\ref{FOAndCIAdlerFunctiond2models}, of which
the upper plot corresponds to FOPT and the lower to CIPT. The dotted,
dot-dashed and double-dot-dashed curves are the 3rd, 4th and 5th order purely
perturbative results respectively. The thick solid line corresponds to the
Borel sum of the reference model. An analogous plot was
already shown as figure~9 in Appendix~B of \cite{BJ2008}. In addition, we now
also display the additional models with $d_2^{\rm PO}=1$ (short-dashed line)
and $d_2^{\rm PO}=-1$ (long-dashed line). Furthermore, the shaded area indicates
the 5th order PT result when the coefficient $c_{5,1}$ is varied in the range
$c_{5,1}=283\pm283$.

The following observations can be made on the basis of
figure~\ref{FOAndCIAdlerFunctiond2models}. For an asymptotic expansion, the
last included term should provide an approximate error estimate for the full
sum. Though this is strictly true only for sign-alternating asymptotic series,
we expect the estimate not to be wildly violated. Employing the shaded area
as such an error estimate, it is seen that in the Euclidean, $\phi\sim \pi$,
the RM of \cite{BJ2008} lies rather close to this region. On the other hand,
the models with $d_2^{\rm PO}=\pm 1$, even in the Euclidean domain where PT
should work well, lie far from 5th order perturbation theory. Furthermore,
moving away from the Euclidean axis, strong oscillations in
$\Re[\widehat D(\phi)]$ are found in those models. It seems rather unlikely to
us that QCD behaves in this way. Turning the argument around and investigating
which values of $d_2^{\rm PO}$ would yield models compatible with the shaded
area, we roughly obtain the range $-0.55<d_2^{\rm PO}<0$. 

To summarise, there are two arguments in favour of the procedure adopted in
ref.~\cite{BJ2008} for the treatment of the polynomial terms in the ansatz of
eq.~(\ref{CMBJ}). First, the fact that $d_1^{\rm PO}$ turns out to be so small
in the central model fit, together with the observed hierarchy
$d_0^{\rm PO}\gg d_1^{\rm PO}$, which leads to a renormalon dominance of the
coefficients at orders as low as $\alpha_s^2$. Second, the unnaturalness of
the Adler function shape along the circle 
when ${\cal O}(1)$ values of $|d_2^{\rm PO}|$ are imposed in the
extended model suggested in ref.~\cite{DGM} that seems to obstruct duality even
in the Euclidean region.

%%%%%%%%%%%   Matching in large-beta0
\subsection[Matching in the large-$\beta_0$ limit]{\boldmath Matching in the large-$\beta_0$ limit}

Another check whether a simple ansatz such as eq.~(\ref{CMBJ}) can work
can be derived from the  large-$\beta_0$ limit. As discussed in
section~\ref{RenormalonModels}, an analytic result for the Borel-transformed
Adler function is available in this limit, eq.~(\ref{BRuk}), and hence the
exact perturbative coefficients $c_{n,1}$ are known to all
orders~\cite{ben93,bro93}. Here, we propose to emulate the matching procedure
performed in QCD in the context of the large-$\beta_0$ approximation. That is,
we make a simple ansatz similar to the RM and fit the parameters of this ansatz
to the low-order $c_{n,1}$ in the large-$\beta_0$ approximation. We then
compare the so-obtained model for the higher-order terms to the exactly known
ones.

In order to implement the matching procedure in the case of the large-$\beta_0$
limit, we adapt the reference model to the present case by using simple and
double poles (instead of branch cuts) for the renormalon singularities. In the
spirit of eq.~(\ref{CMBJ}), the new model can then be written as
\begin{equation}
B[\widehat D](u) \,=\, \frac{d_2^{\rm IR}}{2-u} + \frac{d_3^{\rm IR}}{(3-u)^{1+
\gamma_3}}  + \frac{d_{1}^{\rm UV}}{(1+u)^2} + d_0^{\rm PO} + d_1^{\rm PO}\,u
+ d_2^{\rm PO}\,u^2 \label{BoModelLBeta}.
\end{equation}
When emulating the procedure of ref.~\cite{BJ2008}, 
we set the term $d_2^{\rm PO}=0$.
In large-$\beta_0$, the IR pole at $u=3$ is a double pole and therefore
$\gamma_3=1$. We simulate our ignorance of the structure of this pole 
in full QCD by taking
either $\gamma_3=0$ or $\gamma_3=1$. From eq.~(\ref{BoModelLBeta}), by changing
$\gamma_3$ and the assumptions about $d_2^{\rm PO}$, we define five different
models and perform the matching to the first coefficients of the exact
large-$\beta_0$ Adler function. The models will be denoted by the values
of these parameters as  $M(\gamma_3;d_2^{\rm PO})$. 
In the remainder of this section the
characteristics of these models are discussed, the matching is described, and
we compare with the exact large-$\beta_0$ limit. 

We begin by performing the matching treating the term $d_2^{\rm PO}$ as
suggested in ref.~\cite{BJ2008}, which is equivalent to setting  
$d_2^{\rm PO}=0$. Furthermore we employ $\gamma_3=1$, which gives a 
double pole for the IR singularity at
$u=3$, in agreement with the exact result eq.~(\ref{BRuk}). This model is
referred to as $M(1;0)$. The renormalon residua
are fixed to the coefficients $c_{3,1}$, $c_{4,1}$, and $c_{5,1}$; the two
polynomial terms are found by enforcing the true large-$\beta_0$ 
values of $c_{1,1}$ and
$c_{2,1}$. The results of this model are shown in the second row of
table~\ref{Residues}. As in the full QCD case, the value of $d_1^{\rm PO}$
turns out to be small, and significantly smaller then $d_0^{\rm PO}$,
indicating the renormalon dominance at intermediate orders. Furthermore, the
residua of the renormalon poles are in the ball-park of the true results,
although they have to compensate for the lack of higher order poles.

\begin{table}[!ht]
\begin{center}
\begin{tabular}{l r r r c r c}
\toprule
& $d_2^{\rm IR}$ & $d_3^{\rm IR}$ & $d_{1}^{\rm UV}\qquad$ &
  $d_0^{\rm PO}$& $d_1^{\rm PO}$& $d_2^{\rm PO}$  \\
 \midrule
large-$\beta_0$ &   $17.84$      & $-10.49$     &   $6.68\times 10^{-2}$ &  $\cdots$     &  $\cdots$      &  $\cdots$          \\
$M(1;0)$        &   $16.53$      & $-45.79$    &   $4.10\times 10^{-2}$ &  $-2.90$       & $-0.44$  &  0 (fixed)   \\
$M(0,0)$        &    $8.34$      & $-18.43$    &   $4.46\times 10^{-2}$  & $\phantom{-}2.25$         & $0.27$   &  0 (fixed)    \\
$M(1;{\rm free})$&  $18.85$     & $-55.02$       &   $3.92\times 10^{-2}$&  $-3.03$&$-0.34$ & $5.79\times 10^{-2}$  \\
$M(1,1)$         &   $56.54$      &$-205.27$       &   $9.76\times 10^{-3}$ & $-5.15$        & $1.31$          &  $\phantom{-}1$ (fixed)  \\
$M(1,-1)$         &  $-23.46$     & $113.67$       &   $7.23\times 10^{-2}$&  $-6.52\times 10^{-1}$&$-2.19$   &  $-1$ (fixed) \\
 \bottomrule
\end{tabular}
\caption{Residues of poles and polynomial parameters of the models discussed
in the text. The first row gives the exact result in the large-$\beta_0$
approximation. The models are defined by the values of $\gamma_3$ and
$d_2^{\rm PO}$ in eq.~(\ref{BoModelLBeta}) and denoted  by
$M(\gamma_3;d_2^{\rm PO})$.}
\label{Residues}
\end{center}
\end{table}

\begin{table}[!ht]
\begin{center}{\small
\begin{tabular}{l r r r r r r r}
\toprule
& $c_{6,1}\qquad$ & $c_{7,1}\qquad$ & $c_{8,1}\qquad$ & $c_{9,1}\qquad$ &
  $c_{10,1}\qquad$ &  $c_{11,1}\qquad$ \\
 \midrule
large-$\beta_0$ &  $-1.99\times 10^3$&$9.86\times 10^4$&$-1.08\times 10^6$&$2.78\times 10^7$&$-5.39\times 10^8$& $1.40\times 10^{10}$ \\
$M(1;0)$          & $-2.46\times 10^3$&$1.08\times 10^5$&$-1.30\times 10^6$&$3.28\times 10^7$&$-6.68\times 10^8$&$1.74\times 10^{10}$  \\
$M(0;0)$          & $-3.53\times 10^3$ &$1.08\times 10^5$&$-1.51\times 10^6$&$3.46\times 10^7$&$-7.38\times 10^8$&$1.87\times 10^{10}$\\
$M(1;{\rm free})$  & input$\quad$ &$1.07\times 10^5$ & $-1.21\times 10^6$ & $3.17\times 10^7$& $-6.35\times 10^8$ & $1.67\times 10^{10}$\\
$M(1;1)$        & $5.69\times 10^3$&$8.54\times 10^4$&$2.74\times 10^5$&$1.39\times 10^7$&$-9.09\times 10^7$&$4.95\times 10^{9\phantom{1}}$  \\
$M(1;-1)$         & $-1.06\times 10^4$&$1.30\times 10^5$&$-2.88\times 10^6$&$5.17\times 10^7$&$-1.25\times 10^9$&$2.98\times 10^{10}$\\
  \bottomrule
\end{tabular}
\caption{Higher-order coefficients from the five models described in the text
compared with the exact large-$\beta_0$ results. Models are defined by the values of $\gamma_3$ and
$d_2^{\rm PO}$ in eq.~(\ref{BoModelLBeta}) as 
$M(\gamma_3;d_2^{\rm PO})$.}
\label{LargeCoeff}}
\end{center}
\end{table}

\begin{figure}[!ht]
\begin{center}
\subfigure[Exact large-$\beta_0$]{\includegraphics[width=.49\columnwidth,angle=0]{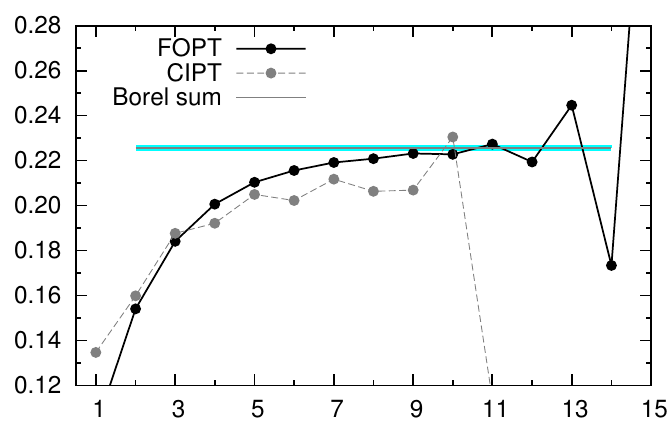}\label{ExactLargeBeta}}
\subfigure[$M(1;0)$]{\includegraphics[width=.49\columnwidth,angle=0]{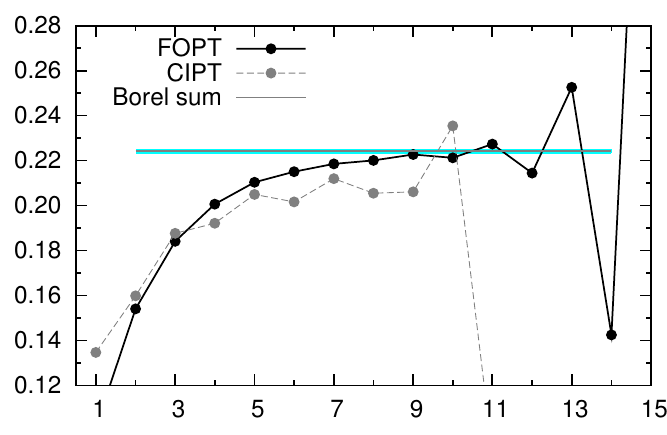}\label{CMFittedLargeBetaPlot}}
\subfigure[$M(0;0)$]{\includegraphics[width=.49\columnwidth,angle=0]{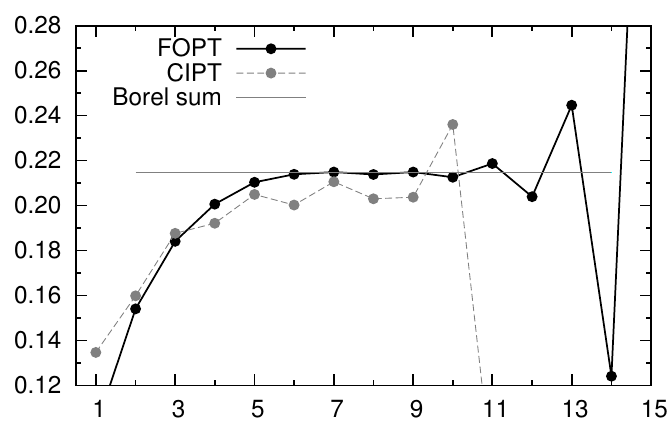}\label{smpIR3Fig}}
\subfigure[$M(1;{\rm free})$]{\includegraphics[width=.49\columnwidth,angle=0]{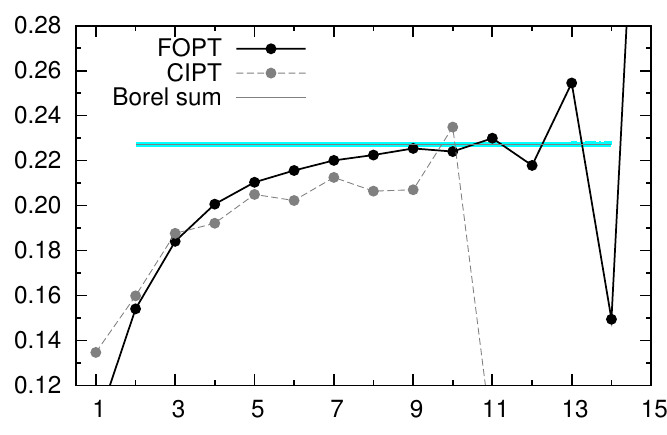}\label{d2POFitLargeBeta0}}
\subfigure[$M(1;1)$]{\includegraphics[width=.49\columnwidth,angle=0]{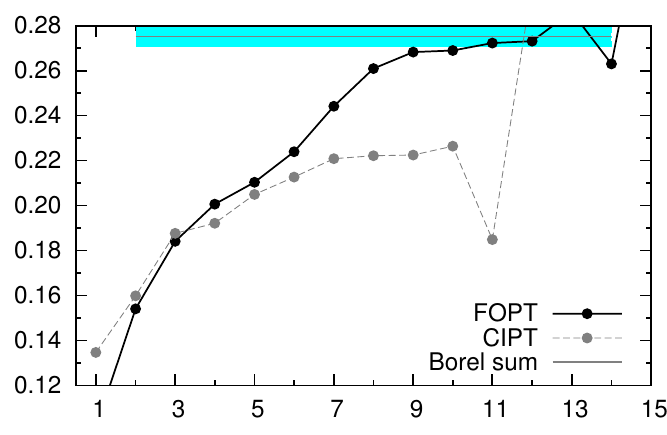}\label{u2FitToLargeBeta}}
\subfigure[$M(1;-1)$]{\includegraphics[width=.49\columnwidth,angle=0]{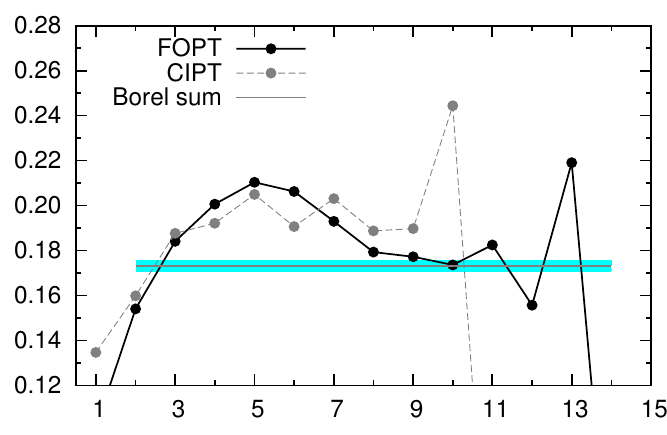}\label{mu2FitToLargeBeta}}
\caption{Values of $\delta^{(0)}_{w_\tau}(m_\tau^2)$ as a function of the order $n$ up
to which the perturbative series has been summed for FOPT (black) and
CIPT (gray). Horizontal gray lines give the Borel resummed result 
 and the bands give the estimated ambiguity. In (a) one
sees the result from the exact large-$\beta_0$ limit. In (b)-(f), we show results
from models matched to the first few coefficients of the
large-$\beta_0$ Adler function (see text).  Models are defined by the values of $\gamma_3$ and
$d_2^{\rm PO}$ in eq.~(\ref{BoModelLBeta}) as $M(\gamma_3;d_2^{\rm PO})$. For consistency  with  large-$\beta_0$, we perform the $\alpha_s$ running at one loop. We use
$\alpha_s(m_{\tau}^2)=0.3186$.} 
\label{CMFitToLargeBeta0_PertSeries}
\end{center}     
\end{figure}

Since the purpose of the model introduced in ref.~\cite{BJ2008} was to decide
upon the best way to perform the RG improvement of the series, it is legitimate
to ask if the description of higher orders is successful in the present case.
This can now be unambiguously tested by comparing the higher order
coefficients predicted by the fitted Adler function with the exact results in
large-$\beta_0$. Numerically, this comparison is shown in
table~\ref{LargeCoeff}. Graphically, FOPT, CIPT, and the resummed results for
the model are shown in figure~\ref{CMFittedLargeBetaPlot} (for $w_\tau$),
whereas the equivalent plot for the exact large-$\beta_0$ results is given in
figure~\ref{ExactLargeBeta}. The qualitative agreement of the results of
table~\ref{LargeCoeff}, together with the striking similarities of
figures~\ref{ExactLargeBeta} and~\ref{CMFittedLargeBetaPlot}, demonstrate that
--- in spite of the simplifications of the model with respect to the exact
results --- the model reproduces faithfully the FOPT and CIPT series up to
higher orders, as well as the Borel resummed value.  

In the previous model, the parameter $\gamma_3$ was fixed to the true value of
the large-$\beta_0$ limit.  Let us investigate the consequences of a wrong
choice for $\gamma_3$. For full QCD this is a relevant open issue as several
operators contribute at $D=6$, and the general renormalon structure has not
yet been established. We define a new model that differs from the previous one
only by having a simple pole at $u=3$, thus $\gamma_3=0$. Accordingly,
we denote this model by $M(0;0)$. Performing the matching,
we find the results of the 3rd row of table~\ref{Residues}. The choice
$\gamma_3=0$ enforces a number of adjustments in the parameters of the model
with respect to the case where $\gamma_3=1$. The residue of the pole at $u=2$
changes and is less well reproduced than in the previous case. The polynomial
terms are different, but the hierarchy between the terms is still preserved.
In particular, 
inspection of the third row of table~\ref{LargeCoeff}, together with
figure~\ref{smpIR3Fig} shows that the higher-order behaviour of FOPT and CIPT
continues to resemble very much the exact result. In agreement with
figure~\ref{ExactLargeBeta}, FOPT approaches the Borel resummed value better
than CIPT. Still, one must admit that there is a shift in the Borel resummed
value of $\sim 5\%$, which is beyond the ambiguity of the exact result
($\sim 1\%$). Nevertheless, the conclusions about FOPT versus CIPT --- and
the superiority of the former --- remain intact.

We now investigate the question of adding a term $d_2^{\rm PO}u^2$ in
eq.~(\ref{BoModelLBeta}). Since all coefficients $c_{n,1}$ of the Adler
function are available in the large-$\beta_0$ approximation, 
we have the freedom to include the coefficient
$c_{6,1}$ in the matching. This allows us to keep the six parameters of the
model free, including $d_2^{\rm PO}$, and then follow a strategy similar to
the one above (we use $\gamma_3=1$). This model is termed $M(1;{\rm
  free})$. We employ the coefficients $c_{4,1}$, $c_{5,1}$, and
$c_{6,1}$ to fix the residua of the renormalon poles, while the polynomial
terms are fixed by $c_{3,1}$, $c_{2,1}$, and $c_{1,1}$. The results after
matching are found in the third row of table~\ref{Residues}. The inclusion of
the free parameter $d_2^{\rm PO}$ preserves the hierarchy
$d_0^{\rm PO}\gg d_1^{\rm PO} \gg d_2^{\rm PO} $, indicating the renormalon
dominance at intermediate and higher orders. This is in agreement with what
was found before, where we kept $d_2^{\rm PO}=0$ fixed. Table~\ref{LargeCoeff}
and Fig~\ref{d2POFitLargeBeta0} show that the quality of the description of
higher orders remains impressive when we introduce the term $d_2^{\rm PO}$
keeping it free and using the procedure of ref.~\cite{BJ2008} to fix it.

Let us turn now to the consequences of having a fixed  
parameter $d_2^{\rm PO}$.
To make contact with ref.~\cite{DGM}, we study the two cases
$d_2^{\rm PO}\pm 1$ (again $\gamma_3=1$). These configurations
are denoted $M(1,\pm 1)$. In table~\ref{Residues}, one sees
some issues that emanate from this choice. The hierarchy of the polynomial
terms is broken and, what is more, the values of the residua of the IR
renormalon poles are very different from the exact ones. We observe large
values of the residua, leading to strong cancellations between the
contributions of the different IR poles, a feature that was already observed
in full QCD. What is then the most faithful description of the higher orders?
Table~\ref{LargeCoeff} demonstrates that the description with fixed
$d_2^{\rm PO}\sim {\cal O}(1)$ is very poor. 
The large-order coefficients are badly reproduced;
even the sign is wrong in two of them. Figures~\ref{u2FitToLargeBeta}
and~\ref{mu2FitToLargeBeta} show that the perturbative series goes astray,
and the Borel resummed results are very different from the exact one shown
in figure~\ref{ExactLargeBeta}.

This exercise shows that the model is incompatible with a $u^2$ term whose
coefficient is of order unity. In real QCD, only four coefficients of the Adler
function are known exactly. In such a scenario, it seems that the best strategy
is to keep $d_2^{\rm PO}=0$, since we learn in large-$\beta_0$ that fixing this
parameter to an arbitrary value is not a better option. 
A final comment is order.
Figure~\ref{CMFitToLargeBeta0_PertSeries} shows results for the kinematic
moment $w_\tau$ only. We have studied all moments displayed in
table~\ref{tab:ws} also for large-$\beta_0$ and the conclusions regarding
the quality of the description are not altered in other cases.

%%%%%%%%%%%%%%%%%%%%%%%%%%%%%%%%%%%%%%%%%%%%%%%%%%%%%%%%%%%%%%%%%%%%%%%

\begin{boldmath}
\section[\boldmath Consequences for the determination of $\alpha_s$]{\boldmath Consequences for the determination of $\alpha_s$}
\end{boldmath}
\label{Alphas}

Based on the perturbative behaviour of the Adler function under two different
assumptions for the higher-order coefficients, we argued in 
section~\ref{Moments} that some weight functions are more suitable 
for an $\alpha_s$ analysis
from hadronic $\tau$ decays than others.  The aim of this section
is to corroborate these findings  comparing the predicted moments with
experimental values. We want to check the internal consistency of the
 predictions from different moments and study the uncertainties 
associated with them once all the contributions are taken into account
(power corrections and duality violations). Rather than comparing
moment predictions with data for some given value of $\alpha_s$, we
perform this consistency check by determining the values of 
$\alpha_s$ from each individual moment. The spread of $\alpha_s$ 
values and uncertainties then provides the desired information. 
We emphasize that the aim of this exercise is {\em not} a precise 
determination of the strong coupling. 

We consider FESRs for the weight-functions of table~\ref{tab:ws}. 
In the notation introduced in eq.~(\ref{RtauDeltas}), we study
the moments $R_{V+A}^{w_i}(m_\tau^2)$, relying on external input for the power
corrections and for DVs. A more comprehensive analysis should include more
than one moment and treat consistently all the ingredients of the theoretical
description: the perturbative series, power corrections, and duality
violations. In such an analysis, an involved non-linear multi-parameter fit
taking into account all correlations is 
unavoidable (see e.g.~\cite{M7C1,M7C2}). Here our error bars in 
$\alpha_s$ are smaller than the ones of a self-consistent analysis 
due to the fact that we do not perform a multi-parameter fit. 
 
The different theoretical components in the computation of the moments are
treated as follows. As discussed in section~\ref{TheoreticalFramework}, the
perturbative FO and CI series are summed up to order $\alpha_s^5$ using
$c_{5,1}=283$. The uncertainty due to the truncation of the series is estimated
by either taking $c_{5,1}=0$ or $c_{5,1}=566$. A third $\alpha_s$ value is
obtained from the Borel resummed results of the reference model.

The power
corrections in the OPE are taken as an external input. Corrections due to
dimension 4, 6, and 8 are considered. (The mass corrections, $D=2$, are also
taken into account although they can safely be neglected due to the smallness
of the quark masses.) For $D=4$, we use the gluon condensate value 
$\langle a G^2\rangle = (0.012\pm 0.012)\,\mbox{GeV}^4$, and include
in the coefficient function the known $\alpha_s$ corrections with
their logarithms~\cite{PP99}. At dimenstion six, the main contributions
arise from the four-quark condensates, since the coefficient of the
three-gluon condensate vanishes at leading order and all terms proportional
to quark masses can safely be neglected. 
One usually resorts to the vacuum saturation approximation to write these
contributions in terms of squares of the quark condensate~\cite{svz79}, introducing
the parameter $\rho_{V+A}$ to account for deviations from this assumption. The $D=6$ corrections
are then proportional to $\rho_{V+A}\langle \bar q q\rangle^2$. In our estimates we
use $\rho_{V+A}=2\pm 1$~\cite{BJ2008} and $\langle \bar q q \rangle(m_\tau) = -(272\pm 15\,\,\,{\rm MeV})^3$~\cite{jam02}. A crude estimate of $D=8$ is included adding the a term $C_{8,V+A}/s^4$
to the Adler function. To evaluate the impact of this contribution we use $C_{8,V+A}=(0\pm 5)\cdot  10^{-3}$. 
Our estimates for $D=6$ and 8
agree, within uncertainties, with the results of the fits of ref.~\cite{M7C2}. 

A phenomenological estimate of the longitudinal contribution from scalar and
pseudoscalar correlators is included in the spirit of
refs.~\cite{Gamizetal03,Gamizetal04}. The weight-function dependence of the
non-perturbative corrections make some of the moments quite insensitive to
the details of the non-perturbative contributions. However, for other
weight-functions, this is not the case and the final $\alpha_s$ values depend
heavily on the non-perturbative input.

Finally, the contribution from DVs to the moments is computed using the results of
ref.~\cite{M7C2}. The corresponding term takes the form
\begin{equation}
\label{dvterm}
\delta_{w_i,V/A}^{\rm DV}(s_0) \,=\, -\,8\pi^2
\int_{s_0}^{\infty} \frac{ds}{s_0}\, w_i(s)\, \rho_{V/A}^{\rm DV}(s) \,,
\end{equation}
where the DV part of the spectral functions, $\rho_{V/A}^{\rm DV}(s)$, is
parametrised as~\cite{CGP}
\begin{equation}
\rho^{\rm DV}_{V/A}(s) \,=\, \exp\left(-\delta_{V/A}-\gamma_{V/A}s\right)
\sin\left(\alpha_{V/A}+\beta_{V/A}s\right) \,.
\end{equation}
The eight parameters for the description of DVs are  taken from the results
of a  fit to $V$ and $A$ using updated OPAL data shown in table~V of ref.~\cite{M7C2}
(we take the  FOPT fit with $s_{\rm min}=1.5$~GeV$^2$). Their correlations are employed
in the Monte Carlo estimate of the error induced
by DVs. Note, however, that within our simplified $\alpha_s$ determination in which DVs and power corrections are 
not determined self-consistently, the precise  DV parameters are not important for
the following. Within other errors, the DV term (\ref{dvterm}) has a negligible
impact on $\alpha_s$ from the moments discussed below.

The experimental counterparts of the moments $R_{V+A}^{w_i}(m_\tau^2)$
are obtained performing a discretised version of the integral given in
eq.~(\ref{Rtauth}), where the spectral functions are the ones from the ALEPH
collaboration after the 2008 update performed in ref.~\cite{Davier}. It is
known that in this data set contributions to the correlations due to the
unfolding procedure were inadvertently omitted~\cite{Manchester}. At present,
in a precise determination of $\alpha_s$, the safest option is to turn to an
updated version of the OPAL data~\cite{OPAL,M7C2}. Nevertheless, for our
exploratory purposes, this omission is harmless and would only alter the
experimental error bars, that are potentially underestimated.
\label{pageALEPH}

\begin{figure}[!t]
\begin{center}
\includegraphics[width=.93\columnwidth,angle=0]{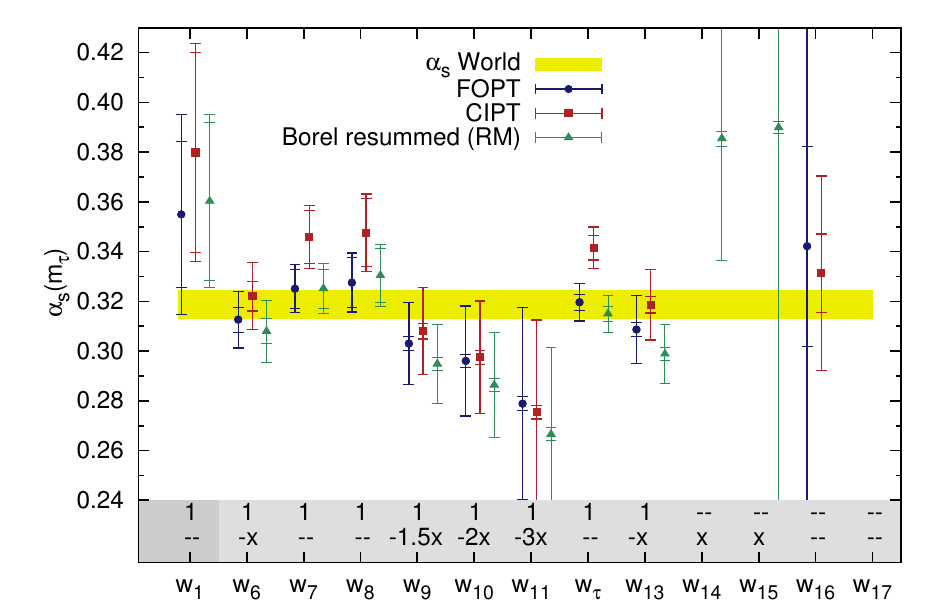}
\caption{Results of $\alpha_s(m_\tau)$ from $V+A$ sum rules constructed with the
weight functions of table~\ref{tab:ws} and using the 2008 version of
ALEPH data ~\cite{Davier}. The absence of a point means that no
reasonable value of $\alpha_s$ was found.  Inner error bars give
solely the experimental error; outer bars include the theory error as
well. In the lower shaded band we show explicitly the constant term  and
the term proportional to $x$ for the weight functions (when present). All the 
moments are pinched except for $w_1$. The world average of $\alpha_s$ is that 
of the PDG~\cite{pdg12}: $\alpha_s(m_\tau)= 0.3186 \pm 0.0056$.}
\label{AlphasPlot}
\end{center}     
\end{figure}

In figure~\ref{AlphasPlot}, we display results of this simplified $\alpha_s$
analysis for 13 of the weight functions of table~\ref{tab:ws}, 
and using FOPT, CIPT, and
the Borel sum within the RM. (We do not show results for the monomials $w_2$
to $w_5$.) The inner error bars give the experimental errors, while the
external error bars include the theoretical error as well. The relative size
of the experimental uncertainties depends strongly on the moment considered.
This is understandable since the spectral functions have larger relative
uncertainties in the higher-energy part. Moments that emphasise this region
(or that do not suppress it) such as $w_1(x)=1$ are penalised and have
significantly larger uncertainties. In general, pinched moments suppress the
edge of the spectrum and have smaller relative uncertainties.

We consider several sources of theoretical errors in $\alpha_s$. The first one
is the truncation of perturbation theory, which is estimated by varying the
coefficient $c_{5,1}=283\pm 283$.  Another source of theoretical error is the
residual renormalisation-scale dependence. To estimate this uncertainty, we
re-express the FO series in terms of a different scale $\mu^2$, rather than
using $a(m_\tau^2)$, and vary this scale. The residual dependence
is estimated in CIPT in a similar fashion by setting the scale to
$\mu^2=-\xi^2s_0x$ in eq.~(\ref{del0}). This generates additional logarithms
that must be taken into account in $\delta^{(0)}_{{\rm CI},w_i}$. In the Borel 
resummed model, this scale/model uncertainty is also estimated by taking $\mu^2=\xi^2m_\tau^2$,
determining $a(\xi m_\tau)$, and evolving the result back to $\mu=m_\tau$. In all cases,
the scale is varied in the interval $0.5\,m_\tau^2 \leq\mu^2 \leq 1.5\, m_\tau^2$.
The uncertainty due to $D=4$ contributions is estimated varying the gluon condensate
in the interval $\langle aG^2\rangle=(0.012\pm 0.012)$ GeV$^{4}$. Uncertainties from $D=6$ and
$D=8$ are computed from the propagation of the error  on the quantities $\rho_{V+A}$, $\langle \bar q q \rangle $, and $C_{8,{V+A}}$.
 Finally, the uncertainty due to
the DV term is estimated from a Monte Carlo sample of parameters generated
according to the results  found in table V of ref.~\cite{M7C2}.
The outer error bars in fig.~\ref{AlphasPlot} contain the sum in quadrature of
all these errors together with the experimental uncertainty.

Let us turn now to an analysis of the results shown in Fig.~\ref{AlphasPlot}. A
first important point is that for the moments considered ideal in the study of $\delta^{(0)}_{w_i}$,
namely, $w_7$, $w_8$, and $w_\tau$, we obtain an $\alpha_s$ value
compatible with the world average within our simplified analysis. The results
from FOPT and the reference model are particularly close to the world average,
while CIPT lies about two sigma away. This is a consequence of the fact that
$\delta^{(0)}_{w_i}$ dominates these moments, and the FOPT perturbative series shows
a good behaviour. They are quite insensitive to the the power corrections
and DVs and, accordingly, have relatively small theoretical uncertainties.
Their experimental uncertainties are  smaller than in other moments due to the
suppression of the edge of the spectral functions.

The situation is radically different for $w_{14}$, $w_{15}$, and $w_{17}$ for
which neither FOPT nor CIPT were able to give a reasonable value of $\alpha_s$.
The Borel resummed result from the reference model of ref.~\cite{BJ2008} does
not yield any acceptable value for $\alpha_s$ for $w_{16}$ and 
$w_{17}$ either. For $w_{14}$ and $w_{15}$ abnormally high values with huge 
uncertainties  are obtained. 
This is a manifestation of the bad convergence properties
observed for the family of moments $w^{(1,k)}$, defined in eq.~(\ref{ALEPHMom})
and used e.g. in refs.~\cite{ALEPH, ALEPH2, OPAL, Davier}, combined with the fact that
these moments receive very large contributions from power corrections
 (of the order of $50\%$ for $w_{17}$). 
 Note that
 with the present treatment of power corrections the theoretical error
 of these moments is by far dominated by perturbative (scale and $c_{5,1}$ variations)
 uncertainties. This corroborates the conclusion of
section~\ref{Moments}, namely, that these moments are not the optimal choice
for an $\alpha_s$ analysis from $\tau$ decays.  For $w_{16}$, FOPT and CIPT yield a value of $\alpha_s$ with large errors dominated by power corrections of 
dimension 6 and 8 with a sizeable contribution from the perturbative series. The
result in FOPT is particularly unstable, leading to  larger uncertainties.

The scenario for the moments that contain an $x$ term and the unity lies in
between the previous two. Moments like $w_6$ and also $w_{13}$ still 
yield reasonable values of $\alpha_s$. On the other hand, moments 
like $w_9$, $w_{10}$, $w_{11}$ give lower values of $\alpha_s$. This 
can be understood since for these moments the $D=4$ correction is 
increasingly important, due to the higher coefficients of the $x$ term. In
our estimate, these $D=4$ corrections are positive which leads to 
lower values of $\alpha_s$. This is not the case in $w_{13}$ due 
to the $D=6$ and $D=8$ contributions arising from the terms  
$-3x^2+5x^3$ in the weight function, that compensate the $D=4$ corrections.
Since these moments are more sensitive to $D=4$, they exhibit
larger theoretical uncertainties arising predominantly from
$\delta^{(4)}_{w_i,V/A}$.

Finally, from $w_1=1$, a value of $\alpha_s$ compatible with the world average
is also obtained. This moment has a much larger experimental uncertainty due
to the lack of pinching and the corresponding higher contribution from the edge
of the spectrum.

A general observation is that the size of the discrepancy between FO and CI
also depends strongly on the moment. Moments such as $w_{9}$, $w_{10}$, and
$w_{11}$ tend to minimise the discrepancy while the kinematical moment
$w_{\tau}$, and to some extent $w_{7}=1-x^2$ and $w_{8}=1-x^3$, tend to
maximise this difference. We have seen that FOPT is a better approximation to
the Borel resummed results in the reference model; therefore, $\alpha_s$ values
from FOPT are systematically closer to the Borel 
resummed ones than CIPT values.\\

%%%%% Conclusions
\section{Conclusions}
\label{Conclusions}

In the first part of this work, we have analysed the moment dependent
features of the perturbative expansion of the Adler function needed in
the theoretical description of hadronic $\tau$ decays. A systematic
study of this moment dependence is important since it serves as a
guide for future analyses of $\alpha_s$, and provides further
insight on the question whether FOPT or CIPT is a better framework
for $\alpha_s$ extractions. To analyse the higher order
behaviour of the series we employed two models. The first is the
reference model of ref.~\cite{BJ2008}, which gives --- we believe ---
a plausible representation of the QCD Adler function. In this
model, FOPT is the prescription that gives the best approximation to
the Borel resummed results. In order to assess possible model
dependencies in our conclusions, we have also employed a 
model where the IR pole at $u=2$, which is the most important 
one for the perturbative behaviour in intermediate orders, 
is artificially suppressed; this description
favours CIPT. The behaviour of $\delta^{(0)}_{w_i}$ for a large
collection of moments can then be divided into a small number of
classes.  The general behaviour of the perturbative series can be
traced back to the singularities of the Borel transformed Adler
function and to simple features of the weight-functions.

We have shown that some moments have better perturbative properties 
than others. Our conclusions are based on the inspection of the
perturbative series and on a simplified $\alpha_s$ extraction from
each one of the moments.  In particular, polynomial pinched
weight-functions that contain the unity and do not contain a term
proportional to $x$ are found to be optimal. This is due to the good
behaviour of the perturbative series in these cases, where at least
one of the methods, FOPT or CIPT, provides a good approximation to the
Borel resummed results. Another positive feature of these moments is
the small contamination by power corrections and DVs. We have shown
that some of the moments used in the literature, such as the family of
moments $w^{(1,k)}$ [see eq.~(\ref{ALEPHMom})], employed e.g. by ALEPH and
OPAL~\cite{ALEPH,OPAL,Davier}, are perturbatively unstable and do not provide
good approximations to the Borel resummed results of the models
investigated. In the best of the cases, FOPT is capable of approaching
the Borel results only at higher orders that are not available in the 
real QCD case. In addition, the ambiguity introduced by the power
corrections is of the same order as the perturbative contribution,
rendering the extraction of $\alpha_s$ and condensates from these
moments unreliable (similar observations were made in
ref.~\cite{MY08}). Moments that contain an $x$ term lie in between
the two latter groups.  In the reference model, where the
contribution of the leading IR pole is significant, they display 
run-away behaviour in the perturbative series,  which engenders 
larger uncertainties and potential instabilities in $\alpha_s$ results. 
The contribution from $D=4$ in the OPE can
also be quite sizeable depending on the coefficient of the $x$ term in the weight
function. This makes the use of these moments for determinations of $\alpha_s$ 
and condensates somewhat problematic, although the
evidence against their use is less compelling than in the case of moments
with only higher powers of~$x$. 

Overall,  regarding the FOPT/CIPT
comparison, the moment analysis confirms the conclusions drawn 
from the inclusive tau hadronic width \cite{BJ2008} in the following 
sense: whenever perturbatively well-behaved moments are considered, 
FOPT shows better behaviour in the reference model that 
is believed to incorporate the main known features of large-order 
behaviour in QCD. CIPT underestimates the resummed value and 
therefore leads to systematically larger $\alpha_s$ values.

In section~\ref{Plausibility}, we have provided further evidence to
the plausibility of the reference model suggested in ref.~\cite{BJ2008}. The
matching procedure with the first two terms of a polynomial in $u$ in
eq.~(\ref{CMBJ}) was justified based on the behaviour of the Adler
function on the complex plane and on comparisons with the
large-$\beta_0$ limit. This limit  provides a laboratory for
renormalon models, since the exact result to all orders is known.
We have shown that models containing solely the leading singularities
capture  the general features of the exact result surprisingly accurately.
What is more, the use of these models is sufficient to
decide upon the best prescription for the RG improvement of the perturbative
series. It therefore appears that 
the reference model is solid and survives criticisms raised in the 
literature.

\acknowledgments\vspace{-0.3cm}
 MB is supported in part by the Gottfried Wilhelm Leibniz programme of
 the {\it Deutsche Forschungsgemeinschaft} (DFG). The work of DB is
 supported by the Alexander von Humboldt Foundation.  MJ is partially
 supported by CICYTFEDER-FPA2008-01430, FPA2011-25948, SGR2009-894,
 and the Spanish Consolider-Ingenio 2010 Program CPAN (CSD2007-00042).

\end{document}